\documentclass[12pt]{article}

\usepackage[fleqn]{amsmath}
\usepackage{amssymb} 
\usepackage{hhline}
\usepackage[scale={.75,.75}]{geometry}
\usepackage[T1]{fontenc}
\usepackage{bbm} 
\usepackage{fleqn} 
\usepackage{mathrsfs} 
\usepackage{longtable}
\usepackage{caption}
\usepackage{cite}
\usepackage{epsfig}

\DeclareMathOperator{\tr}{tr}

\newcommand{\I}{\text{i}}

\newcommand{\E}[1]{\ensuremath{\text{E}_{#1}}} 
\newcommand{\G}[1]{\ensuremath{\text{G}_{#1}}}
\newcommand{\SO}[1]{\ensuremath{\text{SO}\!\left(#1\right)}}
\newcommand{\SU}[1]{\ensuremath{\text{SU}\!\left(#1\right)}}
\newcommand{\U}[1]{\ensuremath{\text{U}\!\left(#1\right)}}
\newcommand{\Z}[1]{\ensuremath{\mathbbm{Z}_{#1}}} 
\newcommand{\jmod}[2]{#1 \, {\rm mod} \, #2}

\newcommand{\T}{\ensuremath{\boldsymbol{10}}}
\newcommand{\F}{\ensuremath{\boldsymbol{5}}}
\newcommand{\Tb}{\ensuremath{\boldsymbol{\overline{10}}}}
\newcommand{\Fb}{\ensuremath{\boldsymbol{\bar{5}}}}

\newcommand{\jvs}{\rule[-7pt]{0.00pt}{20pt}}

\numberwithin{equation}{section}
\numberwithin{table}{section}
\allowdisplaybreaks

\begin{document}
\date{\mbox{ }}

\title{ 
{\normalsize     
DESY 08-075\hfill\mbox{}\\
July 2008\hfill\mbox{}\\}
\vspace{1cm}
\bf Higgs versus Matter\\ 
in the Heterotic Landscape \\[8mm]}
%
\author{W.~Buchm\"uller and J.~Schmidt\\[2mm]
{\normalsize\it Deutsches Elektronen-Synchrotron DESY, Hamburg, Germany}
}
\maketitle

\thispagestyle{empty}

\begin{abstract}
\noindent
In supersymmetric extensions of the standard model there is no basic
difference between Higgs and matter fields, which leads to the well known
problem of potentially large baryon and lepton number violating interactions.
Although these unwanted couplings can be forbidden by continuous or
discrete global symmetries, a theoretical guiding principle for their choice
is missing. We examine this problem for a class of vacua of the
heterotic string compactified on an orbifold. As expected, in general there 
is no difference between Higgs and matter. However, certain vacua happen to 
possess unbroken matter parity and discrete $R$-symmetries which single out 
Higgs fields in the low energy effective field theory. We present a method 
how to identify maximal vacua in which the perturbative contribution to the 
$\mu$-term and the expectation value of the superpotential vanish.
Two vacua are studied in detail, one with two pairs of Higgs doublets 
and one with partial gauge-Higgs unification.
\end{abstract}

\newpage

\section{Introduction}

In the standard model there is a clear distinction between Higgs and matter:
Quarks and leptons are chiral fermions whereas a  scalar field describes the 
Higgs boson. The most general renormalizable lagrangian consistent with
gauge and Lorentz invariance yields a very successful description of
strong and electroweak interactions \cite{rpp06}. Furthermore, with appropriate
coefficients, the unique dimension-5 operator can account for Majorana 
neutrino masses, and the baryon number violating dimension-6 operators are 
consistent with the experimental bounds on proton decay.

In supersymmetric extensions of the standard model the distinction between
Higgs and matter is generically lost. Since the lepton doublets and one of
the Higgs doublets have the same gauge quantum numbers the most general 
supersymmetric gauge invariant lagrangian contains unsuppressed $R$-parity 
violating terms which lead to rapid proton decay. In grand unified models 
(GUTs) \cite{rpp06} colour triplet exchange can also generate dangerous 
baryon number 
violating dimension-5 operators. These problems can be overcome by introducing
continuous or discrete symmetries which distinguish between Higgs and matter 
fields, such as $R$-symmetry, Peccei-Quinn type symmetries or matter parity.
However, in the context of four-dimensional (4D) field theories the origin 
and theoretical justification of these symmetries remain unclear.

Higher-dimensional theories provide a promising framework for unified 
extensions of the supersymmetric standard model \cite{w85}. In particular the
heterotic string \cite{ghx85} with gauge group $\E8\times\E8$ is the
natural candidate for a unified theory including gravity. Its 
compactifications on orbifolds \cite{dhx85,inq87} yield chiral gauge theories 
in four dimensions including the standard model as well as GUT gauge groups.
During the past years some progress has been made in deriving
unified field theories from the heterotic string  
\cite{krz04,fnx04, bhx04}, separating the GUT scale from the string
scale on anisotropic orbifolds \cite{ht04}, and a class of compactifications 
yielding supersymmetric standard models in four dimensions have been 
successfully constructed \cite{bhx05,lnx06,nrr08}.

The heterotic string model \cite{bhx05} has a 6D orbifold GUT limit, where
two compact dimensions are much larger than the other four, with 6D bulk
gauge group $\SU6$ and unbroken $\SU5$ symmetry at two fixed points. The
corresponding supergravity model has been explicitly constructed in 
\cite{bls07}, and it has been shown that all bulk and brane anomalies
are canceled by the Green-Schwarz mechanism. Furthermore,
a class of vacua has been found which have a pair of
bulk Higgs fields and two $\SU5$ bulk families in addition to the two 
$\SU5$ brane families. At the $\SU5$ fixed points these fields form an
$\SU5$ GUT model. In 4D one obtains one quark-lepton `family' and a pair of 
Higgs doublets from split bulk multiplets together with the two brane families.

What distinguishes Higgs from matter fields with the same $\SU5$ quantum 
numbers in an orbifold GUT? In the vacuum studied in \cite{bls07} there is no 
distinction, which leads to unacceptable $R$-parity violating Yukawa couplings.
In \cite{lnx06} interesting 4D vacua with unbroken matter parity were found, 
which allow to forbid the dangerous $R$-parity violating couplings. Some of 
these vacua also have gauge-Higgs unification for which an intriguing 
relationship exists between $\mu$-term and gravitino mass. Indeed, several
vacua with semi-realistic Yukawa couplings could be identified where to order 
six in powers of standard model singlets $\mu$-term and gravitino 
mass both vanish. 

In this paper we further analyse the vacua of the 6D orbifold GUT \cite{bls07}.
Since $M_{\mathrm{GUT}} \ll M_{\mathrm{string}}$, we consider vacua with
expectation values (VEVs) of all 6D zero modes. One then obtains further 
vacua with unbroken matter parity. The localized Fayet-Iliopoulos terms
of anomalous $\U1$ symmetries may indeed stabilize two compact dimensions
at the GUT scale \cite{bls07,bcs08} but the study of stabilization and 
profiles of bulk fields \cite{gno02} is beyond the scope of this paper. 
In the following we concentrate on local properties
of the model at the GUT fixed points, in particular
the decoupling of exotics and the generation of superpotential
terms.

The existence of a matter parity is not sufficient to distinguish Higgs
from matter. One also needs that the $\mu$-term is much smaller than the
decoupling mass of exotic states. In principle, 
there are two obvious solutions: Either a non-zero $\mu$-term is 
generated at very high powers in standard model singlets, or the perturbative 
part of the $\mu$-term vanishes exactly and a non-perturbative 
contribution, possibly related to supersymmetry breaking, yields a
correction of the order of the electroweak scale. In Section~4, we shall
discuss how to identify `maximal' vacua with vanishing $\mu$-term, as well as 
extended vacua with $\mu$-terms generated at high orders. This is the main 
point of our paper.

The maximal vacua with vanishing $\mu$-term do not include the case of
gauge-Higgs unification. Instead, we find a vacuum with two pairs of massless
Higgs doublets and one with partial gauge-Higgs unification only for $H_u$
which gives mass to up-type quarks. This is perfectly consistent with 
the fact that a large top-quark mass is singled out. The original symmetry
between $\F$- and $\Fb$-plets is violated by selecting vacua where matter
belongs to $\Fb$- and $\T$-plets.

There are also other promising approaches which use elements of unification 
to find realistic string vacua. This includes compactifications on 
Calabi-Yau manifolds with vector bundles \cite{bhox05,bd05,bhw05,bmw06,ac07,knw08},
which are related to orbifold constructions whose 
singularities are blown up \cite{lrx06,gkx08}. Very recently, also interesting GUT models
based on F-theory have been discussed \cite{dw08,bhv08,tw08}.

The paper is organized as follows. In Section~2 we recall some symmetry
properties of effective $\SU5$ field theories, which are relevant for the
$\mu$-term and baryon number violating interactions. The relevant features
of the 6D orbifold GUT model \cite{bls07} are briefly reviewed in Section~3.
New vacua of this model with vanishing $\mu$-term and gravitino mass are 
analyzed in Section~4, and the corresponding unbroken discrete 
$R$-symmetries are determined. Yukawa couplings for these vacua are 
calculated in Section~5.

\section{Effective low energy field theory}
\label{sec:nsec2}
The heterotic 6D GUT model \cite{bls07} has local SU(5) invariance
corresponding to Georgi-Glashow unification. Hence,
the superpotential of the corresponding low energy 4D field theory has the 
general form,
\begin{eqnarray}\label{effective}
W &=& \mu H_uH_d + \mu_i H_u\Fb_{(i)} + C_{ij}^{(u)}\T_{(i)}\T_{(j)} H_u 
+ C_{ij}^{(d)}\Fb_{(i)}\T_{(j)}H_d \nonumber\\
&& + C_{ijk}^{(R)}\Fb_{(i)}\T_{(j)}\Fb_{(k)} 
+ C_{ij}^{(L)}\Fb_{(i)}H_u\Fb_{(k)}H_u 
+  C_{ijkl}^{(B)}\T_{(i)}\T_{(j)}\T_{(k)}\Fb_{(l)}\;, 
\end{eqnarray}
where we have included dimension-5 operators.
Here $i,j,...$ denote generation indices, and for simplicity we have kept the 
$\SU5$ notation. Note that the colour triplets contained in the Higgs fields 
$H_u=\F$ and $H_d=\Fb$ are projected out. $\mu_i$ and $C^{(R)}$ yield the well 
known renormalizable baryon (B) and lepton (L) number violating interactions, 
and the coefficients $C^{(L)}$ and $C^{(B)}$ of the dimension-5 operators 
are usually obtained by integrating out states with masses 
$\mathcal{O}(M_\mathrm{GUT})$. In supergravity theories
also the expectation value of the superpotential is important since it 
determines the gravitino mass. One expects
\begin{equation}
\langle W \rangle \sim \mu \sim M_{\mathrm{EW}},
\end{equation}
if the scale $M_{\mathrm{EW}}$ of electroweak symmetry breaking is related to 
supersymmetry breaking.
 
Experimental bounds on the proton lifetime and lepton number violating 
processes imply $\mu_i \ll \mu$, $C^{(R)} \ll 1$ and 
$C^{(B)} \ll 1/M_\mathrm{GUT}$. Furthermore, one has to accommodate the
hierarchy between the electroweak scale and the GUT scale,
$M_{\mathrm{EW}}/M_\mathrm{GUT} = \mathcal{O}(10^{-14})$. On the other hand, 
lepton number violation should not be too much suppressed, since 
$C^{(L)} \sim 1/M_\mathrm{GUT}$ yields the right order of magnitude for 
neutrino masses. 

These phenomenological requirements can be implemented by means of continuous 
or discrete symmetries. Imposing an additional $\U1$ factor with
\begin{eqnarray}
SU(5)\times U(1)_X &\subset& SO(10)\;, \nonumber\\ 
SU(5)\times U(1)_X &\supset& SU(3)\times SU(2)\times U(1)_Y\times U(1)_{B-L}\;,
\end{eqnarray}
where $Y$ denotes the standard model hypercharge, one has 
$\mu_i = C^{(R)} = C^{(L)} = 0$, 
since these operators contain only $B-L$ violating 
terms. On the other hand, $C^{(B)}$ conserves $B-L$ and is therefore not 
affected. The canonical $\U1_X$ charges read
\begin{equation}
t_X(\T) = \frac{1}{5}\;, \quad t_X(\Fb) = -\frac{3}{5}\;, \quad
t_X(H_u) = -\frac{2}{5}\;, \quad t_X(H_d) = \frac{2}{5}\;, \label{eq:QX}
\end{equation} 
with 
\begin{equation}
t_{B-L} = t_X + \frac{4}{5}\ t_Y \;.
\end{equation}
The wanted result, $\mu_i = C^{(R)} = 0$, $C^{(L)} \neq 0$, can be obtained
with a $\Z2^X$ subgroup of $\U1_X$, which contains the `matter parity' 
$P_X$ \cite{drw82},
\begin{equation}
P_X(\T) = P_X(\Fb) = -1\;, \quad  P_X(H_u) = P_X(H_d) = 1 .
\end{equation}
Matter parity, however, does not solve the problem $C^{(B)} \neq 0$, and also 
the hierarchy $M_\mathrm{EW}/M_\mathrm{GUT} \ll 1$ remains unexplained.

In supersymmetric extensions of the standard model, electroweak symmetry
breaking is usually tied to supersymmetry breaking. It is then natural to
have $\mu = \mu_i = 0$ for unbroken supersymmetry. 
One easily verifies that
in this case, for $C^{(R)} = C^{(B)} = 0$, the superpotential aquires a unique 
Peccei-Quinn type $\U1_{PQ}$ symmetry with charges
\begin{equation}\label{u1pq}
t_{PQ}(\T) = \frac{1}{2}\;, \quad t_{PQ}(\Fb) = 1\;, \quad
t_{PQ}(H_u) = -1\;, \quad t_{PQ}(H_d) = -\frac{3}{2}\;,
\end{equation} 
together with an additional $\U1_R$ symmetry with $R$-charges
\begin{equation}\label{U1R}
R(\T) = R(\Fb) = 1\;, \quad R(H_u) = R(H_d) = 0 .
\end{equation} 
Note that the $\U1_R$-symmetry implies the wanted relations
$\mu = \mu_i = C^{(R)} = C^{(B)} = 0$, with $C^{(L)}$  unconstrained.
On the other hand, the Peccei-Quinn symmetry only yields 
$\mu = C^{(R)} = C^{(B)} = 0$. 

The latter relations can also be obtained by imposing only a discrete 
$\Z2^{PQ}$ subgroup with $PQ$-parities
\begin{equation}
P_{PQ}(\T) = P_{PQ}(H_d) = -1\;, \quad  P_{PQ}(\Fb) = P_{PQ}(H_u) = 1\;.
\end{equation}
On the contrary, the familiar $R$-parity, which is preserved by non-zero
gaugino masses,
\begin{equation}
P_R(\T) = P_R(\Fb) = -1\;, \quad P_R(H_u) = P_R(H_d) = 1\;,
\end{equation}
implies $\mu_i = C^{(R)} = 0$, whereas $\mu$, $C^{(L)}$ and $C^{(B)}$ are 
all allowed.   

In summary, the unwanted terms in the lagrangian (\ref{effective}) can be
forbidden by a continuous global $R$-symmetry. Supersymmetry breaking will
also break $\U1_R$ to $R$-parity, which may lead to an $R$-axion. The
dangerous terms $\mu$ and $C^{(B)}$ will then be proportional to the soft 
supersymmetry breaking terms and therefore strongly suppressed. Alternatively,
the unwanted terms in (\ref{effective}) can be forbidden by discrete 
symmetries, such as matter parity, PQ-parity or $R$-parity.

In ordinary 4D GUT models continuous or discrete symmetries can be introduced
by hand. It is interesting to see how protecting global symmetries arise
in higher-dimensional theories. The global $U(1)_R$ symmetry (\ref{U1R}) 
indeed occurs naturally \cite{hn01}, and it has been used in 5D and 6D 
orbifold GUTs \cite{abc03}. However, as we shall see
in the following sections, orbifold compactifications of the heterotic
string single out discrete symmetries, which may or may not commute with
supersymmetry.

\section{Heterotic $\SU6$ model in six dimensions}

Let us now briefly describe the main ingredients of the 6D orbifold GUT model
derived in \cite{bls07}. The starting point is the $\E8\times\E8 $  heterotic  
string propagating in the space-time background
$(X_4 \times Y_2)/\Z2 \times M_4$. Here 
$X_4= (\mathbbm{R}^4/\Lambda_{\G2\times\SU3})/\Z3$, 
$Y_2= (\mathbbm{R}^2/\Lambda_{\SO4})$ 
and $M_4$ represents four-dimensional Minkowski space;
$\mathbbm{R}^4/\Lambda_{\G2\times\SU3}$ and $\mathbbm{R}^2/\Lambda_{\SO4}$
are the tori associated with the root lattices of the Lie groups 
$\G2\times\SU3$ and $\SO4$, respectively. 
By construction the $\Z{6-\mathrm{II}}=\Z3\times\Z2$ twist yielding the
orbifold has $\Z3$ and $\Z2$ subtwists which act trivially on the $\SO4$ and 
the $\SU3$ plane, respectively. As a consequence, the model has bulk fields
living in ten dimensions and fields from twisted sectors, which are confined 
to six or four dimensions.

\begin{figure}[t]
\begin{center}
\includegraphics[height=7cm]{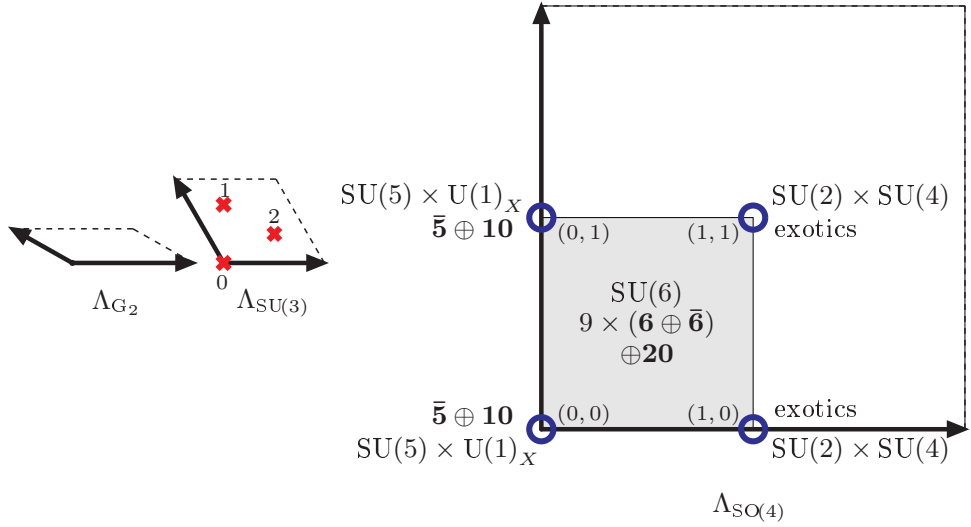}
\caption{\label{fig:6dmodel}The six-dimensional orbifold GUT model with the
unbroken non-Abelian subgroups of the `visible' $\E8$ and the corresponding
non-singlet hyper- and chiral multiplets in the bulk and at the $\SU5$ GUT
fixed points, respectively. Fixed points under the $\Z2$ subtwist in the 
$\SO4$ plane
are labelled by tupels $(n_2,n_2')$, those under the $\Z3$ subtwist in the
$\SU 3$ plane carry the label $n_3=0,1,2$. The $\Z6$ fixed point in the $\G 2$ 
plane is located at the origin.
}
\end{center}
\end{figure}

The model has twelve fixed points\footnote{In the following we shall 
use the terms 'brane' and 'fixed point'
interchangeably. Furthermore, we follow the notations and conventions
of~\cite{bls07}.} where the $\E8\times\E8 $ symmetry is
broken to different subgroups whose intersection is the standard model gauge
group up to $\U1$ factors.
The geometry has an interesting six-dimensional orbifold GUT limit
which is obtained by shrinking the relative size of $X_4$ as compared to $Y_2$.
Such an anisotropy can account geometrically for the hierachy between the 
string scale and the GUT scale.
The space group embedding \cite{bhx05} includes one Wilson line along 
a one-cycle in $X_4$, and a second one as a non-trivial representation of a 
lattice shift within $Y_2$. This leads to the MSSM in the effective 
4D theory \cite{bhx05,lnx06} with the 6D orbifold GUT shown in 
Figure~\ref{fig:6dmodel} as intermediate step \cite{bls07}. At two equivalent
fixed points, labelled as $(n_2,n_2') = (0,0), (0,1)$, the unbroken group 
contains $\SU5$; at the two other fixed points, $(n_2,n_2') = (1,0), (1,1)$,
the unbroken group contains $\SU2\times\SU4$.\footnote{A 5D orbifold GUT model
with the same bulk and brane gauge symmetries and gauge-Higgs unification has 
been constructed in \cite{bn02}; the matter and Higgs sector, however,
is very different from the model \cite{bls07}.}

The 6D orbifold GUT has $\mathcal{N}=2$ supersymmetry and unbroken gauge 
group  
\begin{equation}\label{g6}
{\rm G_6}~=~\SU6\times\U1^3\times \left[\SU3\times\SO8\times\U1^2\right] ,
\end{equation}
with the corresponding massless vector multiplets
\begin{equation}
\left(\boldsymbol{35};1,1\right) + \left(1;\boldsymbol{8},1\right)
+ \left(1;1,\boldsymbol{28}\right) + 5\times \left(1;1,1\right)\;.
\end{equation}
In addition one finds the bulk hypermultiplets
\begin{equation}
\left(\boldsymbol{20};1,1\right) +
\left(1;1,\boldsymbol{8}\right) +
\left(1;1,\boldsymbol{8}_s\right) + \left(1;1,\boldsymbol{8}_c\right) +
4\times \left(1;1,1\right) ,
\end{equation}
where we have dropped the $\U1$ charges. 
It is convenient to decompose all $\mathcal{N}=2$ 6D multiplets in terms of
$\mathcal{N}=1$ 4D 
multiplets. The 6D vector multiplet splits into a pair of 4D vector and chiral
multiplets, $A = (V,\phi)$, whereas a hypermultiplet contains of a pair of 
chiral multiplets, $H = (H_L, H_R)$; here $\phi$ and $H_L$ are left-handed,
$H_R$ is right-handed. It is often convenient to use the charge conjugate field 
$H_R^c$ instead of  $H_R$ so that all degrees of freedom are contained in left-handed 
chiral multiplets. In the following we use the same symbol for a hypermultiplet and its 
left-handed chiral multiplet; the superscript '$c$' indicates that the field is the charge 
conjugate of a right-handed chiral multiplet contained in a hypermultiplet.
As an example,  the chiral multiplets $\F$ and $\Fb^c$ are both $\F$-plets of $\SU5$, 
but they belong to different hypermultiplets which transform as $\F$ and $\Fb$, respectively. 

As we shall see, the four 
non-Abelian singlets, denoted as $U_1...U_4$, play a crucial
role in vacua with unbroken matter parity; the $\SU6$ 
$\boldsymbol{20}$-plet contains part of one quark-lepton generation.
At the $\SU5$ fixed points one has
\begin{equation}\label{bulk}
	{\boldsymbol{35}} = {\boldsymbol{24}} + {\F} + {\Fb} + {\boldsymbol{1}}\;, \quad
	{\boldsymbol{20}} = {\T} + {\Tb}\;.
\end{equation}

In addition to the vector and hypermultiplets from the untwisted sector of
the string, there are 6D bulk fields which originate from the twisted
sectors $T_2$ and $T_4$ of the $\Z{6-\mathrm{II}}$ orbifold model. 
They are localized at the fixed points of the $\Z3$ 
subtwist in the $\SU3$ plane, but bulk fields in the $\SO4$ plane 
which is left invariant by this subtwist. In contrast, fields of the 
twisted sectors $T_1, T_5$ and $T_3$ are localized at fixed points in the 
$\SO4$ plane. For 
simplicity, we shall list in the following only the states of the `visible' 
sector, the complete set of fields can be found in \cite{bls07}.
For each of the three fixed points in the $\SU3$ plane, one finds
\begin{equation}\label{hyper6}
3 \times (\boldsymbol{6}_{n_3} + \boldsymbol{\bar{6}}_{n_3} 
+ Y_{n_3} + \bar{Y}_{n_3}) , \quad n_3 = 0,1,2 ,
\end{equation}
where the omitted $U(1)$ charges depend on $n_3$. The multiplicity factor 3 is
related to three different localizations in the $\G2$ plane; 
$Y_{n_3}$ and $\bar Y_{n_3}$ denote singlets under the non-Abelian part of 
$G_6$. At the $\SU5$ fixed points $n_2=0$, Eq.~(\ref{hyper6}) reads
\begin{equation}\label{hyper5}
3 \times (\boldsymbol{5}_{n_3} + \boldsymbol{\bar{5}}_{n_3} 
+ X_{n_3} + \bar{X}_{n_3} + Y_{n_3} + \bar{Y}_{n_3})\;, \quad n_3 = 0,1,2\;,
\end{equation}
where $X_{n_3}, \bar X_{n_3}$ denote $\SU5$ singlets.
Note that each $\mathcal{N}=2$ hypermultiplet $H$ contains two $\mathcal{N}=1$ 
chiral multiplets $H$ and $H^c$ with opposite gauge quantum numbers.

At the two inequivalent fixed points in the $\SO4$ plane the bulk gauge
group $G_6$ is broken to the subgroups $G_{n_2=0}$ and $G_{n_2=1}$,
respectively,
\begin{align}
	G_{n_2=0} &= \SU5 \times \U1^4 \times \left[ \SU3 \times \SO8
			\times \U1^2 \right] , \\	
	G_{n_2=1} &= \SU2 \times \SU4 \times \U1^4 \times \left[ \SU2' \times \SU4'
			\times \U1^4 \right] . 
\end{align}
At these fixed points
 $\mathcal{N}=1$ chiral multiplets
from the twisted sectors $T_1/T_5$ and $T_3$ are localized. At each $\SU5$
fixed point one has
\begin{equation}\label{local}
\Fb + \T + N^c + S_1 + \ldots + S_8\;.
\end{equation}
This provides two quark-lepton families and additional singlets whose vacuum
expectation values, together with those of $X_{n_3}$ and $Y_{n_3}$ can
break unwanted $\U1$ symmetries. Note that $\Fb$, $\T$ and $N^c$ form together
a $\boldsymbol{16}$-plet of $\SO{10}$ which is unbroken at two equivalent 
fixed points
of the 6D orbifold $T^6/\Z{6-\mathrm{II}}$~\cite{bhx05}. Hence $N^c$ is 
one of the `right-handed' neutrinos in the theory.
 
According to Eqs.~(\ref{bulk}) and (\ref{hyper5}), the 6D theory dimensionally 
reduced to 4D is vectorlike. In terms of $\mathcal{N}=1$ 
chiral multiplets there are two $\T$'s, two $\Tb$'s, 19 $\F$'s and 19 $\Fb$'s.
The chiral spectrum in 4D is a consequence of the further orbifold 
compactification.
At the fixed points of the $\SO4$ plane two chiral families, $\Fb + \T$,
occur. Furthermore, the boundary conditions for the 6D bulk fields at the 
fixed points lead to a chiral massless spectrum. Zero modes require positive
`parities' for bulk fields at all fixed points. As shown in \cite{bls07},
positive parities at the $\SU5$ fixed points
reduce the 18 $\Fb$'s and 18 $\F$'s in Eq.~(\ref{hyper5}) to 
10 $\Fb$'s and 8 $\F$'s, i.e., to a chiral spectrum.   

The model clearly has a huge vacuum degeneracy. In most vacua the standard
model gauge group will be broken. This can be avoided by allowing only
VEVs of the SM singlet fields,
\begin{equation}\label{vacua}
U_1^c,..., U_4, X_0,..., \bar{X}_2^c, Y_0,..., \bar{Y}_2^c,
S_1,..., S_8\;, \label{eq:sings}
\end{equation}  
but most vacua will have a massless spectrum different from the MSSM.
An interesting subset of vacua can be identified by observing that
the products $\F_{n_3}\F^c_{n_3}$ and $\Fb_{n_3}\Fb_{n_3}^c$ are total gauge 
singlets for which one can easily generate masses at the $\SU5$ fixed points.
This allows the decoupling of 6 pairs of $\F$'s and $\Fb$'s \cite{bls07},
\begin{equation}\label{pairwise}
  W \supset M_* \left(\F_0\F_0^c + \Fb_0\Fb_0^c + \F_1\F_1^c 
  + \Fb_1\Fb_1^c + \F_2\F_2^c + \Fb_2\Fb_2^c \right)\;,
\end{equation} 
after which one is left with three $\F$-plets, five $\Fb$-plets and two 
$\T$-plets, 
\begin{equation}\label{allbulk}
\F,\; \Fb,\; \F^c_0,\; \Fb^c_0,\; \F_1,\; \Fb_1,\; 
\F^c_2,\; \Fb_2;\;\; \T,\; \Tb^c\;.
\end{equation}
The decoupling scale $M_*$ will be discussed in more detail later on.
We are now getting rather close to the standard model. The bulk fields,
together with the localized fields (\ref{local}), can account for four 
quark-lepton families, and the additional three pairs of $\F$- and 
$\Fb$-plets may contain a pair of Higgs fields. 

How can one distinguish between Higgs and matter fields and which fields
should be decoupled? The discussion in Section~2 suggests to search for the 
$\U1_X$ symmetry among the six $\U1$ factors at the $\SU5$ fixed points,
so that the extended $\SU5\times\U1_X$ gauge symmetry contains $\U1_{B-L}$,
\begin{equation}
t_X = \sum_{i=1}^5 a_i t_i + a_6 t_6^0 , \quad
t_{B-L} = t_X + \frac{4}{5} t_Y .
\end{equation}  
Here $t_1, \dots, t_6^0$ are generators of the six local $\U1$ 
factors\footnote{Note that the $t_i$ are orthogonal but
not normalized, $t_i \cdot t_j = {\rm diag}(1,1,6,1,3,30)$, 
where $t_6 \equiv t_6^0$.} at 
$n_2=0$ (cf.~\cite{bls07}), and $t_Y$ is the hypercharge generator in $\SU5$. 
For completeness all charges of the remaining $\SU5$ multiplets and 
the singlets (\ref{eq:sings}) are listed
in Tables~\ref{tab:multiplets} and \ref{tab:singlets}, respectively.

We can now  demand the 
canonical $\U1_X$ charges (\ref{eq:QX}) for the localized fields and the bulk
$\T$- and $\Tb^c$-plets. This fixes four coefficients:
$a_1 = a_2 = 2a_4, a_3 = -1/3, a_6 = 1/(15)$. Two $\F$- and two $\Fb$-plets
then have the charges of the Higgs multiplets $H_u$ and $H_d$, respectively,
\begin{equation}
t_X (\F) = t_X (\Fb^c_0) = -\frac{2}{5} , 
\quad t_X (\Fb) = t_X (\F^c_0) = \frac{2}{5} .
\end{equation}
This leaves $\Fb_1$, $\F^c_2$ and $\Fb_2$ as candiates for matter fields. The
requirement to identify two $\Fb$-plets which, together with $\T$ and $\Tb^c$,
form two generations, uniquely determines the last two coefficents,
$a_1 = 1$ and $a_5 = 1/6$, so that 
\begin{equation}
t_X = t_1 + t_2 -\frac{1}{3}t_3 + \frac{1}{2}t_4 + \frac{1}{6}t_5 
+ \frac{1}{15} t_6^0 .
\end{equation}  
The remaining charge assignments read
\begin{equation}
t_X (\F_1) = - t_X (\Fb_1) = -\frac{2}{5}\;, \quad
t_X (\Fb_2) = t_X (\F^c_2) = -\frac{3}{5}\;.
\end{equation}
One can also embed the $\U1_{PQ}$ symmetry (\ref{u1pq}) in the product 
$\U1^6$. One finds
\begin{equation}
t_{PQ} =  -\frac{1}{2}\left(t_1 + t_2\right) + \frac{1}{6}t_3 - 
\frac{1}{2}t_4 + \frac{1}{6}t_5 +  \frac{1}{15} t_6^0 .
\end{equation}  
However, in the vacua considered in the next section, this symmetry is
completely broken.

\begin{table}[t]
  \begin{center}
  \begin{tabular}{|c|ccc|ccccc|}\hline
   & $\F$ & $\Fb_0^c$ & $\F_1$& $\Fb$& $\F_0^c$ & $\Fb_1$& $\Fb_2$     &
     $\F_2^c$\jvs\\\hline\hline
     $\U1_X$ & $-\frac{2}{5}$ & $-\frac{2}{5}$ & $-\frac{2}{5}$ & $\frac{2}{5}$
      & $\frac{2}{5}$ & $-\frac{3}{5}$ & $\frac{2}{5}$ &  $-\frac{3}{5}$
      \jvs\\\hline 
      $\SU3\times \SU2$ & $\left(1,\mathbf{2}\right)$ & 
      $\left(\mathbf{3},1\right)$ & $\left(1,\mathbf{2}\right)$ & 
      $\left(1,\mathbf{2}\right)$ & $\left(\bar{\mathbf{3}},1\right)$ &
      $\left(1,\mathbf{2}\right)$ & $\left(\bar{\mathbf{3}},1\right)$ & 
      $\left(1,\mathbf{2}\right) $ \jvs\\\hline
      $\U1_{B-L}$ & 0 & $-\frac{2}{3}$ & 0 & 0 & $\frac{2}{3}$ &  0 & 
      $ -\frac{1}{3}$ & $-1$ \jvs\\\hline 
      MSSM  & $H_u?$ & &$H_u?$ &$H_d?$ & & $H_d?$ & $d_3$ & $l_3$ \jvs\\\hline 
    \end{tabular}
    \caption{$\SU5$ non-singlet chiral multiplets at $n_2=0$. 
	$\SU3\times\SU2$ representations, $B-L$ charges and MSSM 
    identification refer to the zero modes. \label{tab:5decopling}}
  \end{center}
\end{table}
To proceed further we now consider the zero modes of the $\F$- and $\Fb$-plets 
listed in Table~\ref{tab:5decopling}: $\Fb^c_0$ and $\F^c_0$
yield exotic colour triplets and therefore have to be decoupled, 
\begin{equation}\label{exotic}
W \supset M_*'\ \Fb^c_0\F^c_0\;.
\end{equation} 
where the decoupling scale $M_*'$ will be discussed in more detail later on.
$\Fb_2$ and $\F^c_2$ contain a canonical colour-triplet and lepton doublet,
respectively. Finally, $\F$ and $\F_1$ are candidates for $H_u$, whereas 
$\Fb$ and $\Fb_1$ are candidates for $H_d$.

For the matter fields we now have a clear picture. There are two localized
brane families\footnote{Note that subscripts without brackets denote the 
localization of $T_2/T_4$ twisted fields, $n_3 = 0,1,2$. Subscripts with brackets,
(1) \ldots (4), label the four brane and bulk families defined in (3.17) and (3.18).},
\begin{equation}
(n_2,n_2') = (0,0):\ \Fb_{(1)}, \T_{(1)},\quad   
(n_2,n_2') = (0,1):\ \Fb_{(2)}, \T_{(2)},
\end{equation}
and two further families of bulk fields,
\begin{equation}
\Fb_{(3)} \equiv \F^c_2,\;  \T_{(3)} \equiv \T; \quad 
\Fb_{(4)} \equiv \Fb_2,\;  \T_{(4)} \equiv \Tb^c\;. 
\end{equation}
At the fixed points $n_2 = 0$, these chiral $\mathcal{N} = 1$ multiplets
form a local $\SU5\times\U1_X$ GUT theory. The corresponding Yukawa couplings 
are $4\times 4$ matrices which are generated locally \cite{bls07},
\begin{equation}\label{yuk5}
  W_\mathrm{Yuk} =  
  C_{ij}^{(u)}\T_{(i)}\T_{(j)} H_u + C_{ij}^{(d)}\Fb_{(i)}\T_{(j)}H_d ,
\end{equation}
according to the string selection rules. Projecting the bulk fields to their
zero modes,
\begin{align}
\Tb^c: ({\bf 3},{\bf 2}) &=q, \quad \T: ({\bf \bar 3},1) =u^c,\ (1,1)=e^c,& 
\nonumber \\
\Fb_2: ({\bf \bar{3}},1) &=d^c, \quad \F^c_2: (1,{\bf 2}) =l ,&
\end{align}
yields  one quark-lepton generation in the effective 4D
theory. From~(\ref{yuk5}) one deduces the corresponding $3 \times 3$ 
Yukawa matrices, 
 \begin{equation}
  W_\mathrm{Yuk} = Y_{ij}^{(u)} u_i^c q_j H_u + Y_{ij}^{(d)} d_i^c q_j H_d +  
  Y_{ij}^{(l)} l_i e_j^c H_d ,
 \end{equation}
which avoid the unsuccessful $\SU5$ prediction of 4D GUTs.

Like all $\U1$ factors at the $\SU5$ fixed points,
the $\U1_X$ symmetry has to be spontaneously 
broken at low energies. As we saw in Section~2, it is then crucial to 
maintain a $\Z2$ subgroup, which includes matter parity, to distinguish
between Higgs and matter fields. In order to see whether this is possible
in the present model one has to examine the $\U1_X$ charges of the singlet
fields (\ref{vacua}), which are listed in Table~\ref{tab:singlets}. In the 
vacuum selected in \cite{bls07} fields with $t_X = \pm 1$ obtained a VEV 
breaking $\U1_X$ completely. This led to phenomenologically unacceptable 
$R$-parity violating couplings.

Varying the discrete Wilson line in the $\SO4$ plane, in \cite{lnx06} 4D
models with conserved matter parity were found. In these models only SM
singlets with even $B-L$ charge aquire VEVs. These fields are zero modes
of the 4D theory. In a 6D orbifold GUT model, in principle all 6D zero modes
can aquire VEVs, even if they do not contain 4D zero modes, since the negative
mass squared induced by the local Fayet-Iliopoulos terms can compensate
the positive Kaluza-Klein GUT mass term.
Hence, one can include the fields $U_2$ and $U_4$, which have $t_{B-L}=\pm 2$ 
(see Table~\ref{tab:singlets}), in the set of vacuum fields. Not allowing 
VEVs of singlets with $t_{B-L}=\pm 1$ then preserves matter parity. Note that
not all vacua of the 6D orbifold GUT can be obtained from the 4D zero modes.

The pairwise decoupling (\ref{pairwise}), the decoupling of the exotic
$\F$- and $\Fb$-plets, and the matter parity preserving breaking of $\U1_{B-L}$
can be achieved with the minimal vacuum  
\begin{equation}
\mathcal{S}_0=\left\{X_0, \bar X_0^c, U_2, U_4,  S_2, S_5 \right\} .
	\label{eq:S0}
\end{equation}
For the decoupling masses in Eqs.~(\ref{pairwise}) and (\ref{exotic}) one
obtains,
\begin{equation}
M_* = \langle \bar X_0^c S_2 S_5\rangle\;, \quad
M_*' = \langle X_0^c S_2 S_5\rangle\;. \label{eq:Mstars}
\end{equation}
As we shall discuss in detail in the following section, the couplings 
needed to decouple the $\F\Fb$-pairs satisfy all string selection rules.
Note that no exotic matter is located at the fixed points $n_2=0$. 
Most of the exotic matter at $n_2=1$ can be decoupled by VEVs of just a few 
singlet fields (cf.~\cite{bls07}). This decoupling takes place locally at 
one of the fixed points, which is a crucial difference compared to previous 
discussions of decoupling in four dimensions \cite{bhx05,lnx06}. The
unification of gauge couplings yields important constraints on the
decoupling masses $M_*$ and the GUT scale $M_{\mathrm{GUT}}$. 
This question goes beyond the scope of our paper.  
Detailed studies have recently been carried out for the 6D
model \cite{abc03} in \cite{lee06} and for
a heterotic 6D model similar to the one described here in \cite{drw08}.
\begin{table}[t]
	\begin{center}
		\small
			\begin{tabular}{|c||c|c|c|c|c|c||c|c|c|c|c||c|c|c|}
			\hline
\jvs			Multiplet & $t_1$  & $t_2$ & $t_3$ & $t_4$ & $t_5$ & $t_6^0$ &$R_1$&$R_2$&$R_3$ & $k$ & $k n_3$ 
		& $t_X$& $\tilde R_1$	&$\tilde R_2$   \\
			\hline
			\hline
\jvs			 $\T$  &   $-\frac12$ & $\frac12$ & 0 & 0 & 0 & 3
			 & $-1$ & 0&0&0&0&$\frac 15$& $-1$ &$\frac{1}{10}$ \\
\jvs			 $\bar{\T}^c$ & $\frac12$ & $-\frac12$ & 0 & 0 & 0& 3
			& 0 &$-1$&0&0&0&$\frac15$& $-1$ &$\frac{1}{10}$ \\
\jvs			 $\F$& 0&0&0&0&0&$-6$
			&0&0&$-1$&0&0& $-\frac25$&0&$\frac{4}{5}$\\
\jvs			 $\Fb$ &  0&0&0&0&0&6
			&0&0&$-1$&0&0& $\frac25$&0&$\frac{6}{5}$\\
			\hline
\jvs			 $\T_{(1)}, \T_{(2)}$ & 0 & $-\frac16$ & $-\frac12$ & $\frac13$ & 0& $\frac12$ 
			& $-\frac16$&$-\frac13$&$-\frac12$&1&0 & $\frac15$& $-1$&$\frac{1}{10}$\\
\jvs			 $\Fb_{(1)}, \Fb_{(2)}$  & 0 & $-\frac16$ & $\frac32$ & $\frac13$ & 0  & $-\frac32$
			& $-\frac16$&$-\frac13$&$-\frac12$&1&0& $-\frac35$ & $1$&$\frac{7}{10}$\\
			\hline
\jvs			 $\F_0^c$ & 0 & $\frac13$ & $-1$ & $-\frac23$ & 0& 1
			& $-\frac23$&$-\frac13$&0&4&0& $\frac25$  & $1$&$\frac{1}{5}$\\
\jvs			$\Fb_0^c$ & 0 & $\frac13$ & 1 &$-\frac23$ &0& $-1$
			& $-\frac23$&$-\frac13$&0&4&0& $-\frac25$&0&$\frac{4}{5}$\\
\jvs			 $\F_1$ & 0 & $-\frac13$ & $-1$ & $-\frac13$ & $-1$&  $-1$ 
			& $-\frac13$&$-\frac23$&0&2&2& $-\frac25$& $0$&$\frac{9}{5}$ \\
\jvs			 $\Fb_1$& $\frac12$ & $\frac16$ & 0 & $-\frac13$ & $-1$& $1$ 
			& $-\frac13$&$-\frac23$&0&2&2& $\frac25$&0&$\frac{6}{5}$\\
\jvs			 $\F_2^c$  & $-\frac12$ & $-\frac16$ & 0 & $\frac13$ & $-1$& 1
			& $-\frac23$&$-\frac13$&0&4&8 & $-\frac35$& $1$&$-\frac{3}{10}$\\
\jvs			 $\bar{\F}_2$& 0 & $-\frac13$ & 1 &$-\frac13$ & $1$& $1$ 
			& $-\frac13$&$-\frac23$&0&2&4& $-\frac35$& $-1$&$-\frac{3}{10}$ \\
			\hline
		\end{tabular}
		\normalsize
\caption{$\SU5$ non-singlet chiral multiplets at $n_2=0$. The subscripts 
$(1)$ and $(2)$ denote localization at $n_2'=0$ and $n_2'=1$, respectively.
The charges $\frac 12 t_X$ and $\tilde R_2$ agree $\jmod{ }{1}$.}
		\label{tab:multiplets}
	\end{center}
\end{table}
\begin{table}
	\begin{center}
		\small
		\begin{tabular}{|c||c|c|c|c|c|c||c|c|c|c|c||c|}
			\hline
			Singlet & $t_1$  & $t_2$ & $t_3$ & $t_4$ & $t_5$ & $t_6^0$ &$R_1$&$R_2$&$R_3$ & $k$ & $k n_3$ & $t_X$\\
			\hline
			\hline
			 $U_1^c$ & $-\frac12$ & $-\frac12$ & $-3$ & 0 & 0 &0
			&0&$-1$&0&0&0&0\\
			 $U_2$ & $\frac12$ & $\frac12$ & $-3$ & 0 & 0 & 0
			&$-1$&0&0&0&0&2\\
			 $U_3$& $1$ & $-1$ & 0 & 0 & 0 &0
			&$-1$&0&0&0&0&0\\
			 $U_4$& $-1$ & $-1$ & 0 & 0 & 0 & 0
			&$-1$&0&0&0&0&$-2$\\
			\hline
			 $S_1, S_1'$  & $-\frac12$ & $-\frac23$ & $\frac12$ & $\frac13$ & 0& $\frac52$
			&$\frac56$&$-\frac13$&$-\frac12$&1&0 &$-1$\\
			 $S_2, S_2'$  & $\frac12$ & $-\frac23$ & $-\frac12$ & $\frac13$ & 0& $-\frac52$
			&$\frac56$&$-\frac13$&$-\frac12$&1&0&0 \\
			 $S_3, S_3'$ & $\frac12$ & $\frac13$ & $\frac12$ & $\frac13$ & 0 &  $\frac52$
			&$-\frac16$&$\frac23$&$-\frac12$&1&0&1\\
			 $S_4, S_4'$ & $\frac12$ & $\frac13$ & $\frac12$ & $\frac13$ & 0 &  $\frac52$
			&$\frac{11}{6}$&$-\frac13$&$-\frac12$&1&0&1\\
			 $S_5, S_5'$ & $-\frac12$ & $\frac13$ & $-\frac12$ & $\frac13$ & 0&  $-\frac52$
			&$-\frac16$&$\frac23$&$-\frac12$&1&0&0\\
			 $S_6, S_6'$ & $-\frac12$ & $\frac13$ & $-\frac12$ & $\frac13$ & 0&  $-\frac52$
			&$\frac{11}{6}$&$-\frac13$&$-\frac12$&1&0&0\\			
			 $S_7, S_7'$ & 0 & $-\frac16$ & $-\frac12$ & $-\frac23$ & 1& $\frac52$
			&$\frac56$&$-\frac13$&$-\frac12$&1&1&0 \\
			 $S_8, S_8'$& 0 & $-\frac16$ & $\frac32$ & $-\frac23$ & $-1$ & $\frac52$ 
			&$-\frac16$&$-\frac13$&$-\frac12$&1&2&$-1$\\
			\hline
			 $X_0$& 0 & $-\frac13$ & $1$ & $\frac23$ & 0& $5$ 
			&$-\frac13$&$-\frac23$&0&2&0&0\\
			 $X_0^c$& 0 & $\frac13$ & $-1$ & $-\frac23$ & 0& $-5$ 
			&$-\frac23$&$-\frac13$&0&4&0&0\\
			 $\bar X_0$ & 0 & $-\frac13$ & $-1$ & $\frac23$ & 0 & $-5$
			&$-\frac13$&$-\frac23$&0&2&0&0\\
			 $\bar X_0^c$ & 0 & $\frac13$ & $1$ & $-\frac23$ & 0 & $5$
			&$-\frac23$&$-\frac13$&0&4&0&0\\
			 $X_1$ & 0 & $-\frac13$ & $-1$ & $-\frac13$ & $-1$& $5$
			&$-\frac13$&$-\frac23$&0&2&2&0\\
			 $X_1^c$ & 0 & $\frac13$ & 1 & $\frac13$ & 1& $-5$
			&$-\frac23$&$-\frac13$&0&4&4&0\\
			 $\bar X_1$ & $\frac12$ & $\frac16$ & 0 & $-\frac13$ & $-1$& $-5$
			&$-\frac13$&$-\frac23$&0&2&2&0\\
			 $\bar X_1^c$ & $-\frac12$ & $-\frac16$ & 0 & $\frac13$ & 1& 5
			&$-\frac23$&$-\frac13$&0&4&4&0\\
			$X_2$ & $\frac12$ & $\frac16$ & 0 & $-\frac13$ & $1$& $5$
			&$-\frac13$&$-\frac23$&0&2&4&1\\
			$X_2^c$ & $-\frac12$ & $-\frac16$ & 0 & $\frac13$ & $-1$& $-5$
			&$-\frac23$&$-\frac13$&0&4&8&$-1$\\	
			$\bar X_2$& 0 & $-\frac13$ & 1 & $-\frac13$ & $1$& $-5$ 
			&$-\frac13$&$-\frac23$&0&2&4&$-1$\\
			$\bar X_2^c$& 0 & $\frac13$ & $-1$ & $\frac13$ & $-1$& $5$ 
			&$-\frac23$&$-\frac13$&0&4&8&1\\
			\hline
			$Y_0$ & 1 & $-\frac13$ & 0 & $\frac23$  & 0& 0
			&$-\frac13$&$-\frac23$&0&2&0&1\\
			$Y_0^c$ & $-1$ & $\frac13$ & 0 & $-\frac23$  & 0& 0
			&$-\frac23$&$-\frac13$&0&4&0&$-1$\\
			$\bar Y_0$ & $-1$ & $-\frac13$ & 0 & $\frac23$ & 0& 0
			&$-\frac13$&$-\frac23$&0&2&0&$-1$\\
			$\bar Y_0^c$ & $1$ & $\frac13$ & 0 & $-\frac23$ & 0& 0
			&$-\frac23$&$-\frac13$&0&4&0&1\\
			$Y_1$  & 0 & $\frac23$ & $-2$ & $-\frac13$ & $-1$& 0
			&$-\frac13$&$-\frac23$&0&2&2&1\\
			$Y_1^c$  & 0 & $-\frac23$ & $2$ & $\frac13$ & $1$& 0
			&$-\frac23$&$-\frac13$&0&4&4&$-1$\\
			$\bar Y_1$ & $\frac12$ & $-\frac56$ & $1$ & $-\frac13$ & $-1$& 0
			&$-\frac13$&$-\frac23$&0&2&2&$-1$\\
			$\bar Y_1^c$ & $-\frac12$ & $\frac56$ & $-1$ & $\frac13$ & $1$& 0
			&$-\frac23$&$-\frac13$&0&4&4&1\\
			$Y_2$ & 0 & $\frac23$ & $2$ & $-\frac13$ & $1$& 0
			&$-\frac13$&$-\frac23$&0&2&4&0\\
			$Y_2^c$ & 0 & $-\frac23$ & $-2$ & $\frac13$ & $-1$& 0
			&$-\frac23$&$-\frac13$&0&4&8&0\\
			$\bar Y_2$ & $\frac12$ & $-\frac56$ & $-1$ & $-\frac13$ & $1$& 0
			&$-\frac13$&$-\frac23$&0&2&4&0\\
			$\bar Y_2^c$ & $-\frac12$ & $\frac56$ & $1$ & $\frac13$ & $-1$& 0
			&$-\frac23$&$-\frac13$&0&4&8&0\\
			\hline
		\end{tabular}
		\normalsize
\caption{Non-Abelian singlets at $n_2=0$. $S_1,...,S_8$ and $S'_1,...,S'_8$ 
are localized at $n_2'=0$ and $n_2'=1$, respectively.}
		\label{tab:singlets}
	\end{center}
\end{table}

The minimal vacuum $\mathcal{S}_0$ has two pairs of Higgs doublets. In order
to have gauge coupling unification, one pair has to be decoupled. This can
be done in various ways by enlarging the minimal vacuum. For the decoupling
the 6D gauge couplings are important. For the bulk fields from the untwisted
sector one has 
\begin{eqnarray}
{\cal L}_H &\supset& \sqrt{2}g\int d^2\theta\ H_R^c(\boldsymbol{20})
H_L(\boldsymbol{20})\phi(\boldsymbol{35})\ +\ \text{h.c.} \nonumber\\
&\supset& \sqrt{2}g\int d^2\theta\ \Tb^c \T\ \F\ +\ \text{h.c.}\;.
\end{eqnarray}
Identifying the $\F$-plet from the gauge multiplet with one Higgs multiplet,
$H_u = \F$, therefore yields the wanted large top-quark Yukawa coupling 
\cite{bhx05,lnx06,bls07}. 

For the Higgs field $H_d$ we shall consider both
options, $H_d = \Fb_1$ and $H_d = \Fb$, to which we refer as partial and
full gauge-Higgs unification, respectively. In the first case, 
the 6D gauge interactions,
\begin{eqnarray}
{\cal L}_H &\supset& \sqrt{2}g\int d^2\theta \left( H_R^c(\boldsymbol{6})
\phi(\boldsymbol{35})H_L(\boldsymbol{6}) + H_R^c(\boldsymbol{\bar{6}})
\phi(\boldsymbol{35})H_L(\boldsymbol{\bar{6}})\right)\ +\ \text{h.c.} 
\nonumber\\
&\supset& \sqrt{2}g\int d^2\theta \left(X_0 \F\F_0^c + \bar{X}_0 \Fb\Fb_0^c + 
X_1^c \F_1\Fb +  \bar{X}_1^c \Fb_1\F + X_2 \F\F_2^c 
+ \bar{X}_2^c \Fb_2\F  \right), 
\end{eqnarray}
can be used to decouple the pair $\Fb\F_1$. The VEV 
$\langle X_1^c\rangle \neq 0$ yields the needed mass term. On the
other hand, $\langle \bar{X}_1^c\rangle = 0$ is required to keep the field 
$\F$ massless. Full gauge-Higgs unification needs  
$\langle X_1^c\rangle = \bar{X}_1^c = 0$. Note that VEVs of $X_0$, $\bar{X}_0$
and $X_2^c$ do not lead to mass terms for zero modes of $\F$ and $\Fb$.

The decoupling terms (\ref{eq:Mstars}) require VEVs of both bulk and localized
fields. Note that the localized singlets $S_2$ and $S_5$ correspond to 
oscillator modes.
As we will see in Section~\ref{sec:examples}, bulk and brane field backgrounds
are typically induced by local FI terms. The non-vanishing 
VEVs of localized fields are related to a resolution of the orbifold 
singularities \cite{lrx06,gkx08}. The study of the blow-up of the 6D
orbifold model to a smooth manifold, and the geometrical 
interpretation of the localized VEVs is beyond the scope of this work.

\section{Vanishing couplings and discrete symmetries}
\label{sec:algos}
The heterotic landscape has a tremendous number of vacua. 
Orbifold compactifications correspond to a subset of vacua with enhanced
symmetries. For `non-standard' embeddings of the space group into the
$\E8\times\E8$ lattice, to which our $\Z{6-\mathrm{II}}$ model belongs,
Fayet-Iliopoulos terms related to anomalous $\U1$'s imply that the orbifold
point in moduli space is a `false vacuum'. In `true vacua' some scalar fields
aquire a non-zero VEV, which spontaneously breaks the large symmetry 
$\G{\mathrm{tot}}$ at the orbifold point to a sbgroup $\G{\mathrm{vac}}$. 
For a given orbifold compactification with typically $\mathcal{O}(100)$ 
massless chiral superfields a huge vacuum degeneracy exists. The identification
of standard model like vacua and their stabilization still is a major problem.

In the 6D orbifold GUT model described in the previous section, we have 
identified fields which provide the building blocks of a local $\SU5$ GUT.
The couplings of the effective field theory are generated by expectation
values of products of $\SU5$ singlet fields. The singlet fields with
non-zero VEVs define a vacuum $\mathcal{S}$ which is restricted by the
requirement that states with exotic quantum numbers are decoupled and 
$\mathcal{N}=1$ supersymmetry is preserved. 
 
The appearence of a coupling between some $\SU5$ non-singlets in the effective 
field theory requires the existence of an operator which involves additional
singlets from the vacuum $\mathcal{S}$.
Such operators are strongly restricted by string selection
rules, which can be expressed as a symmetry $\G{\mathrm{tot}}$ at the
orbifold point. A necessary condition for the absence of a certain coupling 
is then the requirement that for the singlets of the vacuum $\mathcal{S}$ 
the corresponding operators do not exist. The vacuum $\mathcal{S}$
has unbroken symmetry $\G{\mathrm{vac}}\subset\G{\mathrm{tot}}$. Obviously,
a sufficient condition for the absence of a coupling between 
$\SU5$ non-singlets is its non-invariance under $\G{\mathrm{vac}}$.
Both conditions will be studied in the following.

The main question in this section is the absence of unwanted superpotential 
terms in the effective theory. We focus on the $\mu$-term, but the discussion 
can easily be extended to dimension-5 proton decay operators as well as other 
couplings. We shall provide an algorithm for finding `maximal vacua' 
which are `orthogonal' to unwanted terms, and we 
present a method which allows to calculate
vanishing tree-level couplings to all orders in powers of singlets.

\subsection{Orbifold geometry and discrete symmetries}

The geometry of the compact space, its invariance under discrete 
rotations and the localization of fields at fixed points
and fixed planes lead to discrete symmetries \cite{akx08} of the 
superpotential in 4D as well as in 6D at the orbifold fixed points.
The discrete rotations in the $\G2$, $\SU3$ and $\SO4$ planes are associated 
with three $R$-charges $R_i$, $i=1,2,3$,
which are conserved modulo the order $l_i=6,3,2$ 
 of the twist in the respective plane,
\begin{equation}\label{Rrule} 
\sum_{j} R_1^{(j)} = -1\mod6\;,\quad 
\sum_{j} R_2^{(j)} = -1\mod3\;,\quad
\sum_{j} R_3^{(j)} = -1\mod2\;,
\end{equation}
where the sum is over all fields of the particular superpotential term. 

Fields from different twisted sectors $T_k$, $k=1,...,6$ have different 
localization properties. For $k=1,5$ fields are localized at fixed points; $k=2,4$
and $k=3$ correspond to brane fields in the $\SO4$ and $\SU3$ planes,
respectively. For each superpotential term one has 
\begin{equation}\label{krule}
 \sum\limits_{j} k^{(j)} =  0\mod 6\;.
\end{equation}
Furthermore, couplings of fields localized in the $\SU3$ and $\SO4$ planes
have to satisfy the constraints
\begin{eqnarray}
 \text{\SU3\ plane} & : & 
 \sum\limits_{j} k^{(j)} n_3^{(j)} ~=~ 0\mod3\;, \label{trule1}\\
 \text{\SO4\ plane}& : & 
 \sum\limits_{j} k^{(j)} n_2^{(j)}~=~ 0\mod2\;,\quad
  \sum\limits_{j} k^{(j)} n_2^{\prime (j)}~= ~0\mod2\;.\label{trule2}
\end{eqnarray}

The constraints (\ref{Rrule}) - (\ref{trule2}) correspond to
a discrete symmetry which acts on the 6D brane and bulk fields. 
From Tables~\ref{tab:multiplets} and \ref{tab:singlets} one reads off
that $R$-charges of fields from the twisted sector $T_k$ have the form
$R_i[\phi_k] = \jmod{-k/l_i}{1}$.
This implies that the discrete 
rotations 
\begin{align}
	g^{(i)}_m &= e^{2 \pi i \frac{m}{l_i} R_i} \;, \quad m \in \Z{ }\;,
\end{align}
which are of order $l_i^2$, form the group $\Z{l_i}\times\Z{l_i}^{(k)}$.
The group element lies in the latter factor for $m=\jmod{0}{l_i}$.
The superpotential has to transform as
\begin{align}
	g_m^{(i)} W &= e^{-2\pi i \frac{m}{l_i}}W\;, \quad
	m \in \Z{ }\;,
\end{align}
under this product group. For $i=1$ one
deduces that the selection rule (\ref{krule})
is implied by the discrete $R$-symmetries 
(\ref{Rrule}) and not an additional independent condition.

We can make the product structure explicit by expressing the actions
of the two subgroups as
\begin{align}
	\Z{l_i} \;: \quad &h_m^{(i)} =
	e^{2\pi\I \frac{1}{l_i} \left(\jmod{m R_i}{k}\right)}\;, &
	\Z{l_i}^{(k)} \;: \quad &\hat h_{m'}^{(i)} =
	e^{2\pi\I \frac{m'k}{l_i}},
	\quad m,m' \in \mathbbm{Z}\;.
\end{align}
This decomposition applies for all three discrete $R$-symmetries. 
The groups $\Z{3}^{(k)}$ and $\Z{2}^{(k)}$ are subgroups of $\Z{6}^{(k)}$
so that the total $R$-symmetry of the lagrangian is given by
\begin{equation}
\G{R} = \Z6^{R_1} \times \Z3^{R_2} \times \Z2^{R_3} \times \Z6^{(k)}.
\end{equation}

The space selection rules (\ref{trule1}) and (\ref{trule2}) correspond to
further discrete symmetries $\Z3$ and $\Z2$, respectively, which commute
with supersymmetry. One then obtains for the full discrete symmetry,
\begin{equation}
\G{\mathrm{discrete}} = 
\left[\Z6^{R_1} \times \Z3^{R_2} \times \Z2^{R_3} \times \Z6^{(k)}\right]_R 
\times \Z3^{k n_3}\times \Z2^{k n_2} \times \Z2^{k n_2 \prime}\;.
\label{symdisc}
\end{equation}
Introducing the `discrete charge vector' 
\begin{equation}
\mathcal{K}=(R_1, R_2, R_3, k, k  n_3, k  n_2, k  n_2 '),
\end{equation}
all superpotential terms have to obey 
\begin{equation}\label{covW}
\mathcal{K}(W) =\mathcal{K}_{\rm vac},
\end{equation}
where the `discrete vacuum charges' are given by
\begin{equation}
\mathcal{K}_{\rm vac} = \left( \jmod{-1}{6}, \jmod{-1}{3}, \jmod{-1}{2},
\jmod{0}{6}, \jmod{0}{3}, \jmod{0}{2}, \jmod{0}{2} \right) . \label{eq:kvac}
\end{equation}
Covariance of the superpotential $W$ corresponds to invariance
of the lagrangian $W|_{\theta\theta}$.
Together with the gauge symmetry
\begin{equation}
\G{\mathrm{gauge}} =
\SU5 \times \U1^4 \times \left[ \SU3 \times \SO8 \times \U1^2 \right] ,  
\end{equation}
the full symmetry at the $\SU5$ fixed points of the 6D orbifold GUT is
\begin{equation}
\G{\mathrm{tot}} = \G{\mathrm{gauge}} \times \G{\mathrm{discrete}} .
\end{equation}
Defining for the $\U1$ symmetries the charge vector
\begin{equation}
Q = (t_1,...,t_6^0) ,
\end{equation}
gauge invariance of the superpotential implies
\begin{equation}
Q(W) = (0, 0, 0, 0, 0, 0) . \label{gaugeW} \\ 
\end{equation}


Localized FI-terms, related to anomalous $\U1$'s, lead to nonvanishing VEVs
of some 6D brane and bulk fields. This breaks the symmetry of the 6D theory
spontaneously,
\begin{equation}
\G{\mathrm{tot}} \rightarrow \G{\mathrm{vac}}\;.
\end{equation}
We are interested in vacua which preserve $\SU5$. We therefore devide all
fields into two sets, $\SU5$ non-singlets $\phi_i$ and $\SU5$ singlets $s_i$.
A set $\mathcal{S}$ of singlets which aquire VEVs, 
\begin{equation}
 \mathcal{S} = \{s_i|\ t_\mathrm{SU(5)}(s_i) = 0, 
\langle s_i \rangle \neq 0\} ,
\end{equation}
defines a vacuum of the 6D orbifold GUT.

\subsection{Maximal vacua for vanishing couplings}

Consider now a vacuum $\mathcal{S}$ and a superpotential term which can lead
to a coupling for the product $\Phi= \prod_j \phi_j^{m_j}$ of $\SU5$ 
non-singlet fields,
\begin{equation}
W = \lambda \Phi \;, \qquad
	\lambda = \prod_i^{N} s_i^{n_i} , \;
	s_i \in \mathcal{S}, \; n_i, N \in \mathbbm{N}\;.\label{eq:Wcoup1}
\end{equation}
The two conditions (\ref{covW}) and (\ref{gaugeW}) 
can be evaluated separately. First, we factorize a part of $\lambda$ which 
transforms non-trivially under gauge transformations by introducing a
`special monomial' $\lambda_s$,
\begin{equation}
	\lambda = \lambda_0  \lambda_s\;, \quad
	 Q(\lambda_s \Phi) = 0 \;, \quad
	 Q(\lambda_0) = 0\;.
\label{eq:lambdas}
\end{equation}
Generically, the set of monomials 
\begin{equation}
{\rm ker} \, Q(\mathcal{S}) 
\equiv	\left\{\lambda_0\ \Big|\,\lambda_0 = \prod_i^{N} s_i^{n_i},\ 
s_i \in \mathcal{S},\ n_i \in \mathbbm{Z},\ Q(\lambda_0)=0 \Big.\right\}
\end{equation}
is a space of dimension larger than one. Note that we allow both $\lambda_0$ 
and $\lambda_s$ to have sub-monomials with negative exponents
$n_i$, in contrast to their product $\lambda$.\footnote{Negative exponents
are allowed in order to promote the set of all possible
exponents of monomials $\{(n_1, \ldots, n_N),N \in \mathbbm N\}$
to a vector space.}
Clearly, results for $\lambda$ cannot depend on the choice of the special 
monomial $\lambda_s$.
Covariance of the superpotential under the discrete symmetries (\ref{symdisc}) requires
\begin{equation}
\mathcal{K}(\lambda_0) = \mathcal{K}_{\rm vac}-\mathcal{K}(\lambda_s \Phi) \;,
	\label{eq:Rscond}
\end{equation}
which defines the subset of monomials in ${\rm ker} \, Q(\mathcal{S})$
yielding a non-vanishing coupling $\lambda$. 

In order to identify vacua where the superpotential term (\ref{eq:Wcoup1}) 
is forbidden we proceed as follows. The elements of 
${\rm ker} \, Q(\mathcal{S})$ are given by the solutions of the equations
\begin{equation}
Q(\lambda_0) = \sum_{i=1}^N n_i Q(s_i) = 0
\end{equation}
for the charge vector $Q$. The solutions can be represented by vectors
$(n_1,...,n_N)$ which are linear combinations of some basis vectors. These
correspond to basis monomials whose products are the elements of 
${\rm ker} \, Q(\mathcal{S})$.

We now examine the discrete symmetries. After the choice of a special monomial 
$\lambda_s$, Eq.~(\ref{eq:Rscond}) can be evaluated for the basis monomials
of ${\rm ker} \, Q(\mathcal{S})$. Starting from a sufficiently small set 
$\mathcal{S}$ which does not satisfy (\ref{eq:Rscond}), one can subsequently 
add further singlets until a `maximal vacuum' is reached for which the term 
(\ref{eq:Wcoup1}) is forbidden to all orders in powers of singlets. The
generalization of this algorithm to the case of more than one forbidden 
coupling is straightforward. 

\subsubsection{Full gauge-Higgs unification}

As a first example, consider the $\mu$-term in the context of full 
gauge-Higgs unification in our model, $H_u = \F$ and $H_d = \Fb$. 
In that case 
\begin{equation}
\Phi \equiv \Phi_{\rm GHU} = H_u H_d = \F\Fb, \quad
Q(\Phi) = 0,\quad \mathcal{K}(\Phi) = 0\;.
\end{equation}
Note that $\Phi$ is a complete singlet. This leads to $\lambda_s=1$ and the condition
\begin{equation}
\mathcal{K}(\lambda_0) = \mathcal{K}_{\rm vac}
\end{equation}
for an allowed $\mu$-term.
Let us now define the vacuum
\begin{equation}
	\mathcal{S}_1 = \mathcal{S}_0 \cup
	\left\{X_1, \bar X_1, Y_2, S_7\right\} ,
	\label{eq:S1}
\end{equation}
where $S_0$ was defined in~(\ref{eq:S0}).
One easily verifies that the dimension of 
${\rm ker} \, Q(\mathcal{S}_1)$ is four. Basis monomials $\Omega_i$ are listed in
Table~\ref{tab:basker1} from which one reads off that it is impossible to 
satisfy $R_1(\Omega_i)=\jmod{-1}{6}$. Hence, the $\mu$-term is absent in the 
vacuum $\mathcal{S}_1$ to all orders in the singlets.

\begin{table}[t]
\begin{center}
  \begin{tabular}{|c|c|r r r|c|c|}
	  \hline
	  Name & Monomial & $R_1$ & $R_2$ & $R_3$ & $k$ & $k  n_3$ \\
	  \hline
	  \hline
	  $\Omega_1$ & $\bar X_0^c S_2 S_5$ & 0 & 0 & $-1$ & 6&0\\
	  \hline
	  $\Omega_2$ & $X_1 Y_2 S_2 S_5$ & 0 & $-1$ & $-1$ & 6&6 \\
	  \hline
	  $\Omega_3$ & $X_0 \bar X_1 S_5 S_7$ & 0 & $-1$ & $-1$ & 6&3 \\
	  \hline
	  $\Omega_4$ & $X_0 \bar X_1 Y_2 U_2 U_4$ & $-3$ & $-2$ & 0 & 6&6\\
	  \hline
  \end{tabular}
  \caption{Basis monomials of ${\rm ker} Q(\mathcal{S}_1)$ and the 
  corresponding discrete charges. All monomials have $k n_2 = k n_2'=0$.}
\label{tab:basker1}
\end{center}
\end{table}

The vacuum $\mathcal{S}_1$ is maximal since adding any further singlet
respecting matter parity leads to a $\mu$-term. This is demonstrated by 
Table~\ref{tab:proof-S1} where for each possible additional singlet the order 
is listed
at which a $\mu$-term appears. It is intriguing that for some vacua a 
$\mu$-term only occurs at very high orders in the singlets.

As discussed in Section~3, there is another candidate for $H_d$ with even 
matter parity, $\Fb_1$ from the twisted sector $T_2$. The vacuum 
$\mathcal{S}_1$ has only full gauge-Higgs unification if the field $\Fb_1$ 
is decoupled by means of a large mass term together with $\F_1$ which also has
even matter parity. 

Using the method described above we can easily study the 
mass term $\Phi = \F_1\Fb_1$. Choosing as special monomial
$\lambda_s=(X_1 \bar X_1)^{-1}$, which has the convenient feature $Q(\lambda_s \F_1\Fb_1) = 0$,
one obtains
\begin{equation}
\mathcal{K}(\lambda_s \F_1\Fb_1) = 0. \label{eq:ls-pghu-R}
\end{equation}
The conditions for the existence of a $\mu$-term then read
\begin{equation}
\mathcal{K}(\lambda_0) = \mathcal{K}_{\rm vac}, \quad
n_{s}(\lambda) \geq 0 , 
	\label{eq:condn}
\end{equation}
where $n_{s}(\lambda)$ is the exponent of the singlet $s \in \mathcal{S}_1$ 
in the monomial $\lambda = \lambda_0 \lambda_s$. The last condition requires 
the appearance of at least one factor of $\Omega_2$, and $\Omega_3$ or 
$\Omega_4$ from Table~\ref{tab:basker1} in the monomial $\lambda_0$.
However, the $R$-charges of these monomials imply that again it is impossible 
to satisfy the first condition (\ref{eq:condn}) for the vacuum $\mathcal{S}_1$.
Hence, also the mass term $\F_1\Fb_1$ vanishes to all orders in the singlets.
Analogously, one easily verifies that the mass terms $\F\Fb_1$ and
$\F_1\Fb$ vanish as well.

\begin{table}[t]
\begin{center}
  \begin{tabular}{|c|c|c||c|c|}
	  \hline 
      Add & Mass term for $\F\Fb$ & Order& Mass term for $\F_1 \Fb_1$ & Order\\
	  \hline
	  $\bar Y_2$ & 
	  $(X_0 \bar X_0^c \bar X_1 \bar Y_2 (S_5)^2 )^2 \Omega_1 \Omega_4$  & 20
	  &$ (X_0)^2 X_1 \bar X_1 (Y_2)^2 (\bar Y_2)^2 (S_5)^4 \Omega_2 \Omega_4$ & 21\\
	  $\bar Y_2^c$ & 
	  $(\bar Y_2^c S_2 S_7)^2 \Omega_1 \Omega_4$  & 14
	  &$X_0 Y_2 (\bar Y_2^c)^2 (S_2)^3 (S_5)^2 (S_7)^3 \Omega_2 \Omega_4$&21\\
	  $U_1^c$ & $(X_0 \bar X_1 Y_2 U_1^c) \Omega_2$  & 8 
	  &$X_0 (Y_2)^2 U_1^c S_2 S_5$&6\\
	  $U_3$ &
	  $(\bar X_0^c U_3 (S_5)^2)^2 \Omega_2 \Omega_4$  & 17
	  &$X_0 \bar X_0^c (Y_2)^2 U_2 (U_3)^2 U_4 (S_5)^4 \Omega_1$&15\\
	  $S_6$ &
	  $(X_1 Y_2 S_2 S_6) \Omega_4$&9 
	  &$X_0 (Y_2)^2 U_2 U_4 S_2 S_6$&7\\
	  \hline
  \end{tabular}
  \caption{Addition of any further field to $\mathcal{S}_1$ generates monomials
  which induce mass terms for $\F \Fb$ and $\F_1 \Fb_1$.
  Shown are lowest order examples. The monomials $\Omega_i$ are defined in table \ref{tab:basker1}.
  Singlets which complete pairs $A^c A$ are not listed, since they allways allow to form 
  mass terms proportional to $\Omega_1 A^c A$.
  We do only consider singlets which conserve matter parity.
  }
\label{tab:proof-S1}
\end{center}
\end{table}

Adding  further singlets to the vacuum $\mathcal{S}_1$ leads to a
non-zero $\F_1\Fb_1$ mass term as demonstrated in
Table~\ref{tab:proof-S1}. The mass terms for $\F\Fb$ and $\F_1\Fb_1$
are roughly of the some order in the singlets. It is intriguing that in
some cases very high orders occur, which could explain the hierarchy
between the electroweak scale and the GUT scale. However, the main result of 
this section is that the vacuum $\mathcal{S}_1$ does not correspond to 
gauge-Higgs unification. Instead, it represents a model with two pairs of 
Higgs doublets. This may be phenomenologically acceptable, but it is 
inconsistent with gauge coupling unification.

\subsubsection{Partial gauge-Higgs unification}

Consider now the case of partial gauge-Higgs unification, $H_u=\F$ and 
$H_d=\Fb_1$,
\begin{equation} 
\Phi \equiv \Phi_{\rm PGHU} = H_u H_d = \F\Fb_1,
\end{equation}
which can be realized with the vacuum
\begin{equation}
\mathcal{S}_2 = \mathcal{S}_0 \cup \left\{X_1^c, \bar X_1, 
Y_2^c, \bar Y_2, U_1^c, U_3, S_6, S_7 \right\}. \label{eq:S2}
\end{equation}
As discussed in Section~3, the $\F_1\Fb$ pair can be decoupled with the
VEV $\langle X_1^c \rangle \neq 0$. For the new vacuum 
${\rm ker} \, Q(\mathcal{S}_2)$ is again easily calculated, it has dimension
eight. A set of basis monomials is listed in Table~\ref{tab:basker2}.	

For partial gauge-Higgs unification the $\mu$-term is the $\F\Fb_1$ mass
term. Choosing as special monomial $\lambda_s = (\bar X_1)^{-1}$, with
$Q(\lambda_s\F\Fb_1) = 0$, one obtains
\begin{equation}\label{Kpghu}
\mathcal{K}(\lambda_s\F\Fb_1) = (0,0,-1,0,0,0,0).
\end{equation}
The conditions for the existence of a $\mu$-term now read
\begin{align}
\mathcal{K}(\lambda_0) &= 
\left( \jmod{-1}{6}, \jmod{-1}{3}, \jmod{0}{2},\jmod{0}{6}, 
\jmod{0}{3}, \jmod{0}{2}, \jmod{0}{2} \right) , \nonumber\\
n_{s}(\lambda) &\geq 0 , 
\label{eq:condn2}
\end{align}
where $n_{s}(\lambda)$ is the exponent of the singlet $s \in \mathcal{S}_2$ 
in the monomial $\lambda = \lambda_0 \lambda_s$. The last condition requires 
the presence of at least one factor of $\Omega_4'$,  $\Omega_5'$,  
$\Omega_6'$, $\Omega_7'$ or $\Omega_8'$. Since all basis monomials
have even $R_1$ charge the first condition (\ref{eq:condn2}) is always 
violated by monomials in ${\rm ker} \, Q(\mathcal{S}_2)$. Hence, the 
$\mu$-term vanishes again to all orders in the singlets.
\begin{table}
\begin{center}
  \begin{tabular}{|c|c|r r r|c|c|}
	  \hline
	  Name&Monomial & $R_1$ & $R_2$ & $R_3$ & $k$ & $k n_3$\\
	  \hline
	  \hline
	  $\Omega_1'$ & $\bar X_0^c S_2 S_5$ & 0 & 0 & $-1$ & 6&0\\
	  \hline
	  $\Omega_2'$ & $\bar X_0^c X_1^c Y_2^c$ & $-2$ & $-1$ & 0 & 12&12 \\
	  \hline
	  $\Omega_3'$ & $\bar X_0^c (S_5)^2 U_3$ & $-2$ & 1 & $-1$ & 6&0 \\
	  \hline
	  $\Omega_4'$ & $X_0 \bar X_1 S_5 S_7$ & 0 & $-1$ & $-1$ & 6&3 \\
	  \hline
	  $\Omega_5'$ & $X_0 \bar X_0^c X_1^c \bar X_1 U_1^c$ & $-2$ & $-3$ & 0 & 12&6 \\
	  \hline
	  $\Omega_6'$ & $X_0 \bar X_0^c \bar X_1 \bar Y_2 (S_5)^2$ & $-2$ & $-1$ & $-1$ & 12&6 \\
	  \hline
	  $\Omega_7'$ & $X_0 \bar X_0^c \bar X_1 \bar Y_2 (S_6)^2$ & $2$ & $-3$ & $-1$ & 12&6 \\
	  \hline
	  $\Omega_8'$ & $X_0 \bar X_0^c X_1^c \bar X_1 U_2 U_4$ & $-4$ & $-2$ & 0 & 12&6 \\
	  \hline
  \end{tabular}
  \caption{Basis monomials of ${\rm ker} Q(\mathcal{S}_2)$ and their discrete charges.
  All monomials have $kn_2=k n_2'=0$.}
\label{tab:basker2}
\end{center}
\end{table}

The vacuum $\mathcal{S}_2$ is also maximal, since the only possibility to 
enlarge it without breaking matter parity is to add singlets $A$ ($A^c$) 
whose $\mathcal{N} = 2$ superpartners $A^c$ ($A$) already belong to 
$\mathcal{S}_2$. One then obtains the $\mu$-term
\begin{equation}
\mu = \lambda_0\lambda_s, \quad \lambda_0 = A A^c (\Omega_5')^3,
\end{equation}
which is of order 16 in the singlets. This power may be sufficiently high to 
provide an explanation of the hierarchy between the electroweak and the GUT 
scale.

\subsubsection{$\mu$-term and gravitino mass}

The method of maximal vacua also allows to relate the existence of different
couplings. In particular, one can show for full and partial gauge-Higgs
unification that the existence of a $\mu$-term and a singlet contribution 
$W_0$ to the superpotential, which determines the gravitino mass 
$m_{3/2} \propto \langle W_0 \rangle$, are equivalent.

For full gauge-Higgs unification the equivalence follows directly from the fact
that $\mu$ and $W_0$ are given by invariant monomials in 
${\rm ker} \, Q(\mathcal{S})$ \cite{lnx06},
\begin{equation}
	\mu \Phi_{\rm GHU}\;\text{allowed} \quad
	\Leftrightarrow \quad
	W_0 = \mu \;\text{allowed} \;. 
\end{equation}

For partial gauge-Higgs unification the condition for a $\mu$-term 
$\mu \equiv \mu_0\lambda_s$ depends on the quantum numbers of the Higgs fields,
\begin{equation}
\mathcal{K}(\mu_0) = 
\mathcal{K}_{\rm vac} - \mathcal{K}(\lambda_s \Phi_{\rm PGHU}) = 	
\mathcal{K}(W_0) - \mathcal{K}(\lambda_s \Phi_{\rm PGHU}) .
\end{equation}
From Eq.~(\ref{Kpghu}) and Table~\ref{tab:basker2} one reads off 
\begin{equation}
\mathcal{K}(\lambda_s \Phi_{\rm PGHU}) = \mathcal{K}(\Omega_1') =
\mathcal{K}((\Omega_4')^3),
\end{equation}
which implies
\begin{align}
&\mu \Phi_{\rm PGHU} = 
\mu_0(\lambda_s\Phi_{\rm PGHU})\; \text{allowed} \quad
	\Rightarrow \quad
	W_0 = \mu_0  \Omega_1' \; \text{allowed} \;, \\
&	W_0 \; \text{allowed} \quad 
	\Rightarrow \quad
	\mu  \Phi_{\rm PGHU} = W_0 (\Omega_4')^3 (\lambda_s\Phi_{\rm PGHU})\; \text{allowed}.
\end{align}
Note that $\Omega_1'=\bar X_0^c S_2 S_5$ is the monomial used for the 
decoupling of $\F\Fb$ pairs in Section~3.

Our analysis demonstrates that the $\mu$-term and the gravitino mass are
closely related, in particular for vacua with full and partial gauge-Higgs
unification.

\subsection{Unbroken symmetries}

In a given vacuum $\mathcal{S}$ the symmetry at the $\SU5$ fixed points
\begin{equation}
\G{\mathrm{tot}} = \G{\mathrm{gauge}} \times \G{\mathrm{discrete}} 
\end{equation}
is spontaneously broken to some subgroup, 
\begin{equation}
\G{\mathrm{tot}} \rightarrow \G{\mathrm{vac}}(\mathcal{S}),
\end{equation}
which can be identified in the standard manner. Knowledge of 
$\G{\mathrm{vac}}(\mathcal{S})$
is obviously very valuable since it restricts possible terms in the 
superpotential. Forbidden couplings for Yukawa matrices correspond
to `texture zeros'.

Consider a singlet $s_i \in \mathcal{S}$. Under the symmetry $G_{\mathrm{tot}}$
it transforms as
\begin{equation}
s_i \rightarrow e^{2\pi\I\left(\alpha\cdot Q + r\cdot \mathcal{K}\right)} s_i\;.
\end{equation}
Here the vectors $\alpha$ and $r$,
\begin{equation}
\alpha = \left(\alpha_1,...,\alpha_6\right),\; \alpha_i \in \mathbbm{R}, \quad
r = \left(\frac{r_1}{6}, \frac{r_2}{3}, \frac{r_3}{2}, \frac{r_4}{6},
\frac{r_5}{3}, \frac{r_6}{2}, \frac{r_7}{2}\right),\; r_i \in \mathbbm{Z},
\end{equation} 
parametrize the continuous and discrete symmetries of the theory.

A parametrization of the unbroken group $\G{\mathrm{vac}}(\mathcal{S})$ 
in terms of vectors $\alpha'$ and $r'$ can be found by solving the equations
\begin{equation}\label{vacequations}
s_i = e^{2\pi\I\left(\alpha'\cdot Q + r'\cdot \mathcal{K}\right)} s_i, \quad
\forall\; s_i \in \mathcal{S} .
\end{equation}
Knowing the allowed vectors $\alpha'$ and $r'$, the group  
$\G{\mathrm{vac}}(\mathcal{S})$ can be determined.

One unbroken discrete subgroup in both vacua $\mathcal{S}_1$ and 
$\mathcal{S}_2$ is easily identified since $U_2$ and $U_4$ are the only fields
with non-zero $\U1_X$ charge,
\begin{equation}
t_X(U_2) = - t_X(U_4) = 2.
\end{equation}
The smallest $\U1_X$ charge is $t_X(\T) = 1/5$. Hence, $\U1_X$ is broken
to the discrete subgroup $\Z{10}^X$ with elements 
$g_m^{X} = \exp{(2\pi\I \frac{m}{2}t_X)}$, $m\in \mathbbm{Z}$, which contains 
matter parity,
\begin{equation}
P_X = e^{2\pi\I\left(\frac{5}{2}t_X\right)}.
\end{equation}

The identification of the further unbroken symmetries is more cumbersome.
We find that in both vacua no continuous $\U1$ symmetry survives. Solving
explicitly equations (\ref{vacequations}) we find for the vacuum 
$\mathcal{S}_1$, 
\begin{equation}
G_{\rm vac}(\mathcal{S}_1) = \Z3^{\tilde{R}_1} \times \Z{10}^X.
\end{equation}
The elements of the $\Z3$ $R$-symmetry are 
$\tilde{g}^{(1)}_m = \exp{(2\pi\I \frac{m}{3}\tilde{R}_1)}$, 
$m\in\mathbbm{Z}$, with
\begin{equation}
\tilde{R}_1 = \alpha_1\cdot Q + r_1\cdot \mathcal{K}, \quad
\alpha_1 = \left(\frac{5}{2}, \frac{15}{2}, 0, \frac{5}{2}, -\frac{5}{2}, 
\frac{1}{2}\right), \;
r_1 = \left(5, 0, 0, 0, 0, 0, 0\right).
\end{equation}
The `vacuum $R$-charge' is given by
\begin{equation}
	r_1\cdot \mathcal{K}_{\mathrm{vac}} = \jmod{1}{3}\;.
\end{equation}
The $\tilde{R}_1$ charges of the $\SU5$ non-singlets are listed in Table~\ref{tab:multiplets}.
Note that $\tilde{R}_1$ is embedded in the $R$-symmetry
as well as the $\U1$ symmetries of the theory.
 
Following the same procedure for the vacuum $\mathcal{S}_2$, one obtains
the unbroken group
\begin{equation}
G_{\rm vac}(\mathcal{S}_2) = \Z2^{\tilde{R}_2} \times \Z{10}^X.
\end{equation}
The elements of the $\Z2$ $R$-symmetry are 
$\tilde{g}^{(2)}_m = 
\exp{\left(2\pi\I \frac{1}{2}\left(\jmod{m \tilde{R}_2}{t_X}\right)\right)}$, 
$m\in\mathbbm{Z}$, with
\begin{equation}
\tilde{R}_2 = \alpha_2\cdot Q + r_2\cdot \mathcal{K}, \quad
\alpha_2 = \left(7, 0, -\frac{7}{6}, \frac{35}{4}, \frac{7}{12}, 
-\frac{7}{15}\right), \;
r_2 = \left(7, 0, 0, 0, 0, 0, 0\right),
\end{equation}
and vacuum $R$-charge
\begin{equation}
	r_2\cdot \mathcal{K}_{\mathrm{vac}} = \jmod 12\;. \label{eq:RT2vac}
\end{equation}
$\tilde{R}_2$ is again a non-trivial linear combination of 
$\U1$ and discrete $R$-charges.
The $\tilde{R}_2$-charges of the $\SU5$ non-singlets are listed in Table~\ref{tab:multiplets}.

Once the unbroken subgroups are known one can calculate the corresponding 
zeros of the superpotential. Consider again a term of the form 
(\ref{eq:Wcoup1}), which transforms under the discrete symmetry $\Z{l_i}$,
$l_i=3,2$, generated by $\tilde{R}_i$, with 
$i=1,2$, respectively, as 
\begin{equation}
W = \lambda \Phi \rightarrow \lambda\ 
\tilde g_m^{(i)} g_n^{X}\ \Phi
=  e^{2\pi\I \frac{m}{l_i} r_i\cdot \mathcal{K}_{\mathrm{vac}}}\ W
\;, \quad m,n \in \Z{ }\;.
\end{equation}
We thus obtain as sufficient condition for the appearance of a vanishing coupling,
\begin{equation}
\tilde{R}_i (\Phi) \neq \jmod{r_i\cdot\mathcal{K}_{\mathrm{vac}}\ }{l_i} 
\quad  \vee \quad 
\frac 12 t_X(\Phi) \neq \jmod{0}{10}
\quad
 \Rightarrow \quad \langle\lambda\rangle = 0 . \label{eq:condTTvac}
\end{equation}
Given the $\tilde{R}_i$ charges of the $\SU5$ non-singlet fields $\phi_j$
this condition is easily evaluated.

We can now confirm the result from the previous section that
the vacuum $\mathcal{S}_1$ has two massless Higgs pairs. From Table~\ref{tab:multiplets}
we read off
\begin{align}
\tilde{R}_1(\F\Fb) &= \tilde{R}_1(\F_1\Fb) = \tilde{R}_1(\F\Fb_1) =
\tilde{R}_1(\F_1\Fb_1) = \jmod 03 \nonumber\\
&\neq \jmod 13 = r_1\cdot\mathcal{K}_{\mathrm{vac}} .
\end{align}
Extending the vacuum  $\mathcal{S}_1$ by one of the singlets listed in 
Table~\ref{tab:proof-S1} preserves $\Z{10}^X$ but breaks $\Z3^{\tilde{R}_1}$. As a consequence,
Higgs mass terms are generated.

Likewise we can study the symmetry transformations of the above terms 
in
the vacuum $\mathcal{S}_2$,
\begin{align}
	\tilde{R}_2(\F \Fb)=\tilde{R}_2(\F \Fb_1) &= \jmod{0}{2} \;, \quad
	\tilde{R}_2(\F_1 \Fb)=\tilde{R}_2(\F_1 \Fb_1) = \jmod{1}{2}\;.
\end{align}
Furthermore, all $\Z{10}^X$ charges vanish. Recalling (\ref{eq:RT2vac}),
this shows that the unbroken $R$-symmetry forbids
the generation of mass terms for $\F \Fb$ and $\F \Fb_1$, but 
allows them for the two remaining combinations. Indeed, at lowest order we find
the mass term
\begin{align}
	W &= \langle X_1^c \rangle \, \F_1 (\Fb + \epsilon \, \Fb_1) 
	\;, \quad 
	\epsilon = \langle X_0 \bar X_0^c X_1^c Y_2^c S_6 S_7 \rangle\;.
\end{align}
This shows that $\F_1$ decouples
together with a linear combination of $\Fb$ and $\Fb_1$.
The orthogonal linear combination is the down-type Higgs,
\begin{align}
	H_d &= \Fb_1 - \epsilon \; \Fb \;.
\end{align}
It is interesting that the vacuum $\mathcal{S}_2$ leads to a 
down-type Higgs with  dominant component from a twisted sector.
In contrast, the up-type Higgs $H_u=\F$ is a pure gauge field in six
dimensions, which is the reason for the large top-quark mass.
Compared to the case of full gauge-Higgs unification, where both Higgs fields
arise from the untwisted sector, this induces non-trivial discrete $R$-charges
for the product $H_uH_d$.

The discrete R-symmetries $\tilde{R}_1$ and $\tilde{R}_2$ of  the 
vacua $S_1$ and $S_2$, respectively,  may be anomalous~\cite{drw08}. This question is
important  since in the case of  an anomaly one can expect the  generation of  
$\mu$-term and gravitino mass  by nonperturbative effects. These questions will be
studied elsewhere.

\section{Local Yukawa Couplings}
\label{sec:examples}
In the previous section we have identified two vacua with conserved matter
parity and vanishing $\mu$-terms. The first vacuum $\mathcal{S}_1$ 
corresponds to a model with two pairs of massless Higgs doublets, and thus
without gauge coupling unification. We therefore focus on the second
vacuum $\mathcal{S}_2$ with partial gauge-Higgs unification.

The vacuum $\mathcal{S}_2$ contains the brane fields $S_2,S_5,S_6,S_7$ localized at
$(n_2,n_2') = (0,0)$, to which we now add the fields $S_2',S_5',S_6',S_7'$ at the
equivalent fixed point $(n_2,n_2') = (0,1)$, 
\begin{align}
\mathcal{S}_0 &= 
\left\{X_0, \bar{X}^c_0, U_2, U_4, S_2, S_5, S_2', S_5'\right\}\;, \\
\mathcal{S}_2 &= 
\mathcal{S}_0 \cup 
\left\{X_1^c, \bar X_1, Y_2^c, \bar Y_2, U_1^c, U_3, S_6, S_7, S_6', S_7' \right\}\;.
\end{align}
The Higgs fields are $H_u = \F$ and $H_d \simeq \Fb_1$.
The vacuum $\mathcal{S}_2$ has the following properties:
\begin{itemize}
\item {$\U1_X$ is spontaneously broken to $\Z{10}^X$ containing matter parity,}
\item { all vector-like exotics at $n_2=0$ decouple,}
\item { all $D$-terms at $n_2=0$ vanish locally,} 
\item { the $\mu$-term vanishes to all orders in the singlets,}
\item { $\langle W \rangle$ vanishes to all orders in the singlets.}
\end{itemize}
The remarkable last two features are a consequence of an unbroken discrete
$R$-symmetry. The vacuum $\mathcal{S}_2$ is maximal in the sense that adding 
more singlets either breaks matter parity or generates a $\mu$-term.

Low-energy supersymmetry requires vanishing $F$- and $D$-terms. 
In the 6D theory with localized FI-terms the
corresponding equations have complicated solutions, leading to
non-trivial profiles for bulk fields \cite{gno02}. We do not study the
full problem here but focus on the local conditions at the GUT
fixed points $n_2=0$. We expect that the local VEVs can be extended
to full dynamical solutions in six dimensions.

The $\mathcal N=2$ vector multiplet has three
auxilliary fields $D_1,D_2,D_3$ which form a triplet
under $\SU2_R$ and must all vanish
in the bulk. However, at
the fixed points half of the supersymmetry is broken and
the local $\mathcal N=1$ vector multiplet
has an effective $D$-term $D\equiv- D_3+F_{56}$,
where $F_{56}$ is the associated field strength in the $y^5,y^6$
direction. Thus the local $D$-term cancelation condition 
at $n_2=0$ (cf.~\cite{gno02}),
\begin{align}
	&D_3^a = F^a_{56}=\frac{g M_P^2}{384 \pi^2}\frac{\tr t_a}{|t_a|^2} 
		+ \sum_i q^a_i | s_i |^2 , 	\label{eq:dterm}
\end{align}
where $q^a_i$ is the $\U1_a$ charge of the singlet $s_i$,
has always a solution, even for non-vanishing right-hand-side.
This means that in principle localized
FI-terms
do not necessarily induce singlet VEVs and
the corresponding $\U1$ can remain unbroken. However,
since our model has distinct anomalous $\U1$ factors
at the inequivalent fixed points $n_2=0,1$ and a
non-vanishing net anomalous $\U1$ in 4D \cite{bls07},
its global $D$-flat solution 
cannot be of that kind. We rather expect a mixture of
singlet VEVs and a nontrivial gauge backround $\langle F_{56}^{\rm an} \rangle$.

For non-anomalous 
$\U1$'s the local field strength in (\ref{eq:dterm}) in the vacuum 
$\mathcal{S}_2$ can vanish since each of the singlets appears in one of the gauge
invariant basis monomials $\Omega_i'$ of ${\rm ker} \, Q(\mathcal{S}_2)$ 
(cf.~Table~\ref{tab:basker2}).
At $n_2=0$ the model has an anomalous $\U1_{\mathrm{an}}$~\cite{bls07},  
\begin{align}
&	t^0_{\mathrm{an}}=-4 t_2+5 t_4- t_5 
	+ t_6^0, \qquad \tr t^0_{\rm an}/|t^0_{\rm an}|^2=2. 
\end{align}
In fact, also $\langle F_{56}^{\rm an} \rangle$ can vanish since one 
can form monomials of singlets with negative anomalous charge, which are gauge
invariant otherwise. An example is
\begin{align}
	\bar X_0^c X_1^c (\bar X_1)^2 S_5 S_6 (S_7)^2 , 
\end{align}
which has $q_{\rm an} = - 74/3$.

We note that the extension of the vacuum $S_2$ to
a global solution is not straightforward. As demonstrated in Table~\ref{tab:basker2}, it
does not provide uncharged monomials of bulk fields only, which include $\bar Y_2$
or $U_3$. Thus VEVs of these fields are incompatible with $D_3^a=0$. 
One may reduce the vacuum to $S_2\setminus\{\bar Y_2,U_3\}$,
or incorporate profiles of (partly) odd fields. Here we
restrain our attention to local properties of the vacuum
$S_2$ at the GUT fixed points, leaving the problem of global
solutions to further studies.

The $F$-terms $F_i=\partial W/\partial s_i$ vanish  trivially
for all vacuum fields $s_i \in \mathcal{S}_2$, since they only arise from monomials which
contain at least one other singlet with zero vacuum expectation value. 
Thus only monomials of the form $W=(\prod_i s_i) u$, with $s_i \in \mathcal{S}_2$
and $\langle u \rangle=0$, induce non-trivial $F$-terms. 
For the vacuum $\mathcal{S}_2$ there are six such terms, arising from 
$u \in \left\{X_0^c,\bar X_0,X_1,\bar X_1^c,Y_2,\bar Y_2^c\right\}$. Each of these singlets $u$
has a partner $u^c$ which is contained in $\mathcal{S}_2$ and thus has a
non-vanishing vev. Note that $u$ cannot be a singlet with odd matter parity
since the latter is preserved by $\mathcal{S}_2$.
The relevant part of the superpotential is then given by
\begin{align}
	W= \left(a_{u1}+ a_{u2}(\Omega_1')^2+ a_{u3} (\Omega_2')^3 + \cdots\right) \Omega_1' u^c u\;,
\end{align}
where the $\Omega_i'$ were introduced in Table~\ref{tab:basker2}, and $a_{uj}$ are coefficients
labeling all completely invariant monomials which can be constructed from vacuum singlets.
The $F$-term conditions become
\begin{align}
	F_u &\propto a_{u1}+ a_{u2}(\Omega_1')^2+ a_{u3} (\Omega_2')^3 + \cdots = 0 \;.
\end{align}
We expect the existence of non-trivial solutions, with VEVs of the singlets 
$s_i \in \mathcal{S}_2$ determined by the coefficients $a_{uj}$. Explicit finite order
examples for similar models were discussed in \cite{lnx06}.

In the framework of heterotic orbifold compactifications, all couplings of
$\SU5$ non-singlet fields arise from higher dimensional operators. In the 
vacuum $\mathcal{S}_2$, to lowest order in the singlets, we find  the 
$\SU5$ Yukawa couplings for the two brane and two bulk families,
\begin{align}
C^{(u)}
&= \left( a_{ij} \right)
= \left( \begin{array}{c c c c}
 \tilde s^{4} & \tilde s^{4} & \tilde s^{5} & \tilde s^{5} \\ 
 \tilde s^{4} & \tilde s^{4} & \tilde s^{5} & \tilde s^{5} \\ 
 \tilde s^{5} & \tilde s^{5} & \tilde s^{6} & g \\ 
 \tilde s^{5} & \tilde s^{5} & g & \tilde s^{6} \\ 
\end{array} \right), &
C^{(d)}
&= \left( b_{ij} \right)
= \left( \begin{array}{c c c c}
 0 & 0 & 0 & 0 \\ 
 0 & 0 & 0 & 0 \\ 
 \tilde s^{10} & \tilde s^{10} & \tilde s^{6} & \tilde s^{6} \\ 
 \tilde s^{1} & \tilde s^{1} & \tilde s^{2} & \tilde s^{2} \\ 
\end{array} \right) .
\end{align}
Here $\tilde s^n$ denotes one or more monomial of order $n$. Explicit lowest 
order monomials are given in Tables~\ref{tab:c1} and \ref{tab:c2}.
Note that all vanishing terms are texture zeros which are protected by the 
unbroken discrete $R$-symmetry to arbitrary order.
\begin{table}[t]
\begin{center}
\begin{tabular}{|c|c|c|}
\hline 
Coupling & Order & Monomial\\ 
\hline \hline
$a_{11}$ & 
4 &  $($ $\bar{X}_0^c$ $)^2$  $S_2$  $S_5$ \\ 
$a_{12}
$ & 
4 &  $($ $\bar{X}_0^c$ $)^2$  $S_2'$  $S_5$ \\ 
$a_{13}
$ & 
5 &  $($ $\bar{X}_0^c$ $)^2$  $(S_2)^2$  $S_5$ \\ 
$a_{14}
$ & 
5 &  $($ $\bar{X}_0^c$ $)^2$  $S_2$  $(S_5)^2$ \\ 
$a_{22}$ & 
4 &  $($ $\bar{X}_0^c$ $)^2$  $S_2'$  $S_5'$ \\ 
$a_{23}
$ & 
5 &  $($ $\bar{X}_0^c$ $)^2$  $(S_2')^2$  $S_5'$ \\ 
$a_{24}
$ & 
5 &  $($ $\bar{X}_0^c$ $)^2$  $S_2'$  $(S_5)^2$ \\ 
$a_{33}$ & 
6 &  $($ $\bar{X}_0^c$ $)^2$  $($ $S_2$ $)^3$  $S_5$ \\ 
$a_{34}
$ & 
0 & $g$\\ 
$a_{44}$ & 
6 &  $($ $\bar{X}_0^c$ $)^2$  $S_2$  $($ $S_5$ $)^3$ \\ \hline 
\end{tabular} 
\caption{Examples of lowest order monomials for $C^{(u)}_{ij}=a_{ij}$ in the vacuum $\mathcal{S}_2$.}
\label{tab:c1}
\end{center} 
\end{table}
\begin{table}[t]
\begin{center} 
\begin{tabular}{|c|c|c|}
\hline 
Coupling & Order & Monomial\\ 
\hline \hline
$b_{31}$ & 
10 &  $X_0$  $($ $\bar{X}_0^c$ $)^2$  $($ $X_1^c$ $)^2$  $\bar{X}_1$  $\bar{Y}_2$  $U_2$  $U_4$  $S_5$ \\ 
$b_{32}$ & 
10 &  $X_0$  $($ $\bar{X}_0^c$ $)^2$  $($ $X_1^c$ $)^2$  $\bar{X}_1$  $\bar{Y}_2$  $U_2$  $U_4$  $S_5'$ \\ 
$b_{33}$ & 
6 &  $X_0$  $X_1^c$  $\bar{X}_1$  $\bar{Y}_2$  $S_6$  $S_7$ \\ 
$b_{34}$ & 
6 &  $\bar{X}_0^c$  $($ $X_1^c$ $)^2$  $Y_2^c$  $S_6$  $S_7$ \\ 
$b_{41}$ & 
1 &  $S_5$ \\ 
$b_{42}$ & 
1 &  $S_5'$ \\ 
$b_{43}$ & 
2 &  $S_2$  $S_5$ \\ 
$b_{44}$ & 
2 &  $($ $S_5$ $)^2$ \\ \hline 
\end{tabular} 
\caption{Examples of lowest order monomials for $C^{(d)}_{ij}=b_{ij}$ in the vacuum $\mathcal{S}_2$.}
\label{tab:c2}
\end{center} 
\end{table}
\begin{table}[t]
\begin{center} 
\begin{tabular}{|c|c|c|}
\hline 
Coupling & Order & Monomial\\ 
\hline \hline
$c_{11}$ & 
11 &  $($ $X_0$ $)^2$  $($ $\bar{X}_0^c$ $)^2$  $\bar{X}_1$  $Y_2^c$  $U_2$  $S_5$  $S_6$  $(S_7)^2$ \\ 
$c_{12}$ & 
11 &  $($ $X_0$ $)^2$  $($ $\bar{X}_0^c$ $)^2$  $\bar{X}_1$  $Y_2^c$  $U_2$  $S_5'$  $S_6$  $(S_7)^2$ \\ 
$c_{22}$ & 
11 &  $($ $X_0$ $)^2$  $($ $\bar{X}_0^c$ $)^2$  $\bar{X}_1$  $Y_2^c$  $U_2$  $S_5'$  $S_6'$  $(S_7')^2$   \\ 
$c_{33}$ & 
12 &   $X_0 (\bar X_0^c)^4 (X_1^c)^2 U_1^c U_2 U_3 S_2 S_5 $ \\ 
$c_{34}$ & 
7 &  $($ $X_0$ $)^2$  $\bar{X}_0^c$  $\bar{X}_1$  $U_2$  $S_6$  $S_7$ \\ 
$c_{44}$ & 
11 &  $($ $X_0$ $)^3$  $($ $\bar{X}_0^c$ $)^2$  $($ $\bar{X}_1$ $)^2$  $U_1^c$  $U_2$  $($ $S_6$ $)^2$ \\ \hline 
\end{tabular}
\caption{Examples of lowest order monomials for $C^{(L)}_{ij}=c_{ij}$ in the vacuum $\mathcal{S}_2$.}
\label{tab:c3}
\end{center} 
\end{table}
After orbifold projection to four dimensions the Yukawa couplings for the zero 
modes read
\begin{align}
	Y^{(u)} &= \left( \begin{array}{c c c}
	a_{11} & a_{12} & a_{14} \\
	a_{21} & a_{22} & a_{24} \\
	a_{31} & a_{32} & a_{34} \\
	\end{array} \right) =
	\left( \begin{array}{c c c}
	\tilde s^4 & \tilde s^4  & \tilde s^5  \\
	\tilde s^4  & \tilde s^4  & \tilde s^5 \\
	\tilde s^5 & \tilde s^5 & g \\
	\end{array} \right) \;, \\
	Y^{(d)} &= \left( \begin{array}{c c c}
	b_{11} & b_{12} & b_{14} \\
	b_{21} & b_{22} & b_{24} \\
	b_{41} & b_{42} & b_{44} \\
	\end{array} \right) =
	\left( \begin{array}{c c c}
	0 & 0  & 0  \\
	0  & 0  & 0 \\
	\tilde s^1 & \tilde s^1 & \tilde s^2 \\
	\end{array} \right) \;, \\
	Y^{(l)} &= \left( \begin{array}{c c c}
	b_{11} & b_{12} & b_{13} \\
	b_{21} & b_{22} & b_{23} \\
	b_{31} & b_{32} & b_{33} \\
	\end{array} \right) =
	\left( \begin{array}{c c c}
	0 & 0  & 0  \\
	0  & 0  & 0 \\
	\tilde s^{10} & \tilde s^{10} & \tilde s^6 \\
	\end{array} \right) \;.	
\end{align}
Clearly, these matrices are not fully realistic since 
$m_e = m_\mu = m_d = m_s = 0$. On the other hand, they show the wanted 
hierarchical structure with a large top-quark mass singled out. Unsuccessful
$\SU5$ mass predictions are avoided since the third 4D quark-lepton family is 
a combination of split multiplets from two 6D families. 

Since $\U1_{B-L}$ is broken the model also predicts Majorana neutrinos. 
`Right-handed' neutrinos with $t_{B-L}$ = 1 can be inferred from Table~\ref{tab:singlets}.
Via the seesaw mechanism they generate light neutrino masses. 
We obtain for the coefficients $C^{(L)}$ (cf.~(\ref{effective})) of the corresponding dimension-5
operator, which can be calculated directly,
\begin{align}
	C^{(L)} &= \left( c_{ij} \right) =
	\left( \begin{array}{c c c c}
 \tilde s^{11} & \tilde s^{11} & 0 & 0 \\ 
 \tilde s^{11} & \tilde s^{11} & 0 & 0 \\ 
 0 & 0 & \tilde s^{12} & \tilde s^{7} \\ 
 0 & 0 & \tilde s^{7} & \tilde s^{11} \\ 
\end{array} \right) .
\end{align}
Examples of lowest order monomials are given in Table~\ref{tab:c3}. 
Projection to four dimensions yields for $\SU2$ doublet zero modes
the $3 \times 3$ sub-matrix with $i,j=1,2,3$.

By construction, the $\mu$-term vanishes to all orders in the vacuum $\mathcal{S}_2$
since it is protected by an unbroken discrete $R$-symmetry. However, this
symmetry is not sufficient to forbid dangerous dimension-5 proton decay operators.
This can be seen from the $\tilde R_2$-charges in Table~\ref{tab:multiplets}, e.g.,
\begin{align}
	\tilde R_2 ( \Fb_{(1)} \T_{(1)} \T_{(1)}\T_{(1)}) &= \jmod{1}{2}\;, \quad
	\tilde R_2 ( \mathcal{K}_{\rm vac}) =\jmod{1}{2} \;.
\end{align}
Since these charges agree and the total $\Z{10}^X$ charge vanishes, 
the proton decay term is not forbidden in the superpotential 
(\ref{effective}).
Indeed, we find a lowest order monomial 
at $\mathcal{O}(7)$, $C^{(B)}_{1111}=(\bar{X}_0^c)^2 X_1^c \bar{X}_1 Y_2^c S_6 S_7$.

Note that the methods presented in Section~\ref{sec:algos} allow to design vacua
with vanishing $\mu$-term and dimension-5 proton decay terms to all orders in
the singlets. An example is the vacuum  $\mathcal{S}_0$, leading
to $\mu= C^{(B)}_{ijkl}=0$. However, this vacuum has other  
problems. It is incompatible with local $D$-term cancelation,
has no gauge-coupling unification and vanishing down-type
Yukawa couplings, $C^{(d)}_{ij}=0$. This demonstrates that the various phenomenological
properties of a vacuum are closely interrelated.

In summary, the vacuum $\mathcal{S}_2$ leads to too rapid proton decay,
and also the quark and lepton mass matrices 
are not fully realistic. However, they show the correct qualitative features
of the standard model, and we are optimistic that a systematic scan of the
heterotic `mini-landscape' can lead to phenomenologically more viable models.

\section{Conclusions}\label{sec:Conclusion}

How to distinguish between Higgs and matter is a crucial question in 
supersymmetric extensions of the standard model, in particular in 
compactifications of the heterotic string. We have analyzed this question
for vacua of an anisotropic orbifold compactification which has an
effective 6D supergravity theory as intermediate step between the 
GUT scale and the string scale.

Our main result is that for generic vacua, there is no difference between 
Higgs and matter, as there is nothing special about the standard model gauge
group. However, certain vacua with standard model gauge group and particle
content can possess discrete symmetries which single out Higgs fields.
They are distinguished from matter fields by a matter parity, and a mass term 
allowed by gauge symmetries is forbidden by an elusive discrete $R$-symmetry, 
a remnant of the large symmetry exhibited by the fundamental theory.

We have identified maximal vacua of a heterotic orbifold model with local 
$\SU5$ unification for which the perturbative contribution to the 
$\mu$-term vanishes. Nonperturbative corrections, possibly related to 
supersymmetry breaking, may then have the size of the electroweak scale.
Alternatively, a non-zero $\mu$-term suppressed by high powers of singlet
fields can appear in extensions of the maximal vacua. 

We have also determined the unbroken discrete $R$-symmetries of the maximal 
vacua. They are judiciously embedded into the large symmetry of the theory,
which is a consequence of the large number of singlet fields forming the
vacuum. It is intriguing that the maximal vacua do not include 
gauge-Higgs unification, but rather partial gauge-Higgs unification for
the Higgs field $H_u$ which gives mass to the up-type quarks. The original
symmetry between $\F$- and $\Fb$-plets is broken by selecting vacua
where matter belongs to $\Fb$- and $\T$-plets.

The method developed to find maximal vacua can be applied to all theories
where couplings are generated by higher-dimensional operators. We have 
focussed on the $\mu$-term, but one can also determine maximal vacua for 
several couplings, like the $\mu$-term and dimension-5 proton decay operators.
In addition to the vanishing of some couplings one may  require
the appearance of certain couplings, like Yukawa couplings or Majorana
neutrino masses. 

The features of the standard model imply strong constraints on  
phenomenolocially allowed vacua. Further important restrictions will
follow from supersymmetry breaking and stabilization of the compact dimensions.
Given the finite number of heterotic string vacua one may then hope to identify
some generic features of standard model vacua, which can eventually be 
experimentally tested.

\vspace{1cm}

\noindent
{\bf\large Acknowlegments}\\

We would like to thank A.~Hebecker, C.~L\"udeling, J.~M\"oller, H.~P.~Nilles 
and M.~Ratz for helpful discussions.


\addcontentsline{toc}{section}{References}

\begin{thebibliography}{10}
\raggedright

\bibitem{rpp06}
  W.~M.~Yao {\it et al.}  [Particle Data Group],
  J.\ Phys.\ G {\bf 33} (2006) 1.
  
\bibitem{w85}
  E.~Witten,
  Nucl.\ Phys.\  B {\bf 258} (1985), 75.
    
\bibitem{ghx85}
D.~J. Gross, J.~A. Harvey, E.~J. Martinec and R.~Rohm, Phys. Rev. Lett.
\textbf{54} (1985), 502; Nucl. Phys. \textbf{B256} (1985), 253.

\bibitem{dhx85}
L.~J. Dixon, J.~A. Harvey, C.~Vafa and E.~Witten, Nucl. Phys. \textbf{B261}
  (1985), 678; Nucl. Phys. \textbf{B274} (1986), 285.

\bibitem{inq87}
L.~E. Ib{\'a}{\~n}ez, H.~P. Nilles and F.~Quevedo, Phys. Lett. \textbf{B187}
  (1987), 25;\\
L.~E. Ib{\'a}{\~n}ez, J.~E. Kim, H.~P. Nilles and F.~Quevedo, Phys. Lett.
  \textbf{B191} (1987), 282.

\bibitem{krz04}
T.~Kobayashi, S.~Raby and R.-J. Zhang, Phys. Lett. \textbf{B593} (2004),
  262 [hep-ph/0403065]; Nucl. Phys. \textbf{B704} (2005), 3 
  [hep-ph/0409098].

\bibitem{fnx04}
S.~F{\"o}rste, H.~P. Nilles, P.~K.~S. Vaudrevange and A.~Wingerter, Phys. Rev.
  \textbf{D70} (2004), 106008 [hep-th/0406208].


\bibitem{bhx04}
W.~Buchm{u}ller, K.~Hamaguchi, O.~Lebedev and M.~Ratz, Nucl. Phys.
  \textbf{B712} (2005), 139 [hep-ph/0412318]; [hep-ph/0512326].

\bibitem{ht04}
A.~Hebecker and M.~Trapletti,
  Nucl.\ Phys.\  B {\bf 713} (2005) 173 [hep-th/0411131].

\bibitem{bhx05}
W.~Buchm{u}ller, K.~Hamaguchi, O.~Lebedev and M.~Ratz, Phys. Rev. Lett.
  \textbf{96} (2006), 121602 [hep-ph/0511035]; Nucl. Phys. \textbf{B785}
  (2007) 149 [hep-th/0606187].

\bibitem{lnx06}
  O.~Lebedev, H.~P.~Nilles, S.~Raby, S.~Ramos-Sanchez, M.~Ratz, 
  P.~K.~S.~Vaudrevange and A.~Wingerter,
  Phys.\ Lett.\  B {\bf 645} (2007) 88 [hep-th/0611095];
  Phys.\ Rev.\  D {\bf 77} (2008) 046013 [arXiv:0708.2691 [hep-th]].
  
\bibitem{nrr08}
For a recent review and extensive references, see\\
  H.~P.~Nilles, S.~Ramos-Sanchez, M.~Ratz and P.~K.~S.~Vaudrevange,
  arXiv:0806.3905 [hep-th].

\bibitem{bls07}
  W.~Buchmuller, C.~Ludeling and J.~Schmidt,
  JHEP {\bf 0709} (2007) 113
  [0707.1651 [hep-ph]].

\bibitem{bcs08}
  W.~Buchmuller, R.~Catena and K.~Schmidt-Hoberg,
  arXiv:0803.4501 [hep-ph].

\bibitem{gno02}
  S.~Groot~Nibbelink, H.~P.~Nilles and M.~Olechowski,
  Phys.\ Lett.\  B {\bf 536} (2002) 270
  [arXiv:hep-th/0203055];\\
  H.~M.~Lee, H.~P.~Nilles and M.~Zucker,
  Nucl.\ Phys.\  B {\bf 680} (2004) 177
  [arXiv:hep-th/0309195].


\bibitem{bhox05}
  V.~Braun, Y.~H.~He, B.~A.~Ovrut and T.~Pantev,
  JHEP {\bf 0506} (2005) 039 [arXiv:hep-th/0502155].
  %

\bibitem{bd05}
  V.~Bouchard and R.~Donagi,
  Phys.\ Lett.\  B {\bf 633} (2006) 783
  [arXiv:hep-th/0512149].

\bibitem{bhw05}
  R.~Blumenhagen, G.~Honecker and T.~Weigand,
  JHEP {\bf 0506} (2005) 020
  [arXiv:hep-th/0504232].

\bibitem{bmw06}
  R.~Blumenhagen, S.~Moster and T.~Weigand,
  Nucl.\ Phys.\  B {\bf 751} (2006) 186
  [arXiv:hep-th/0603015].

\bibitem{ac07}
  B.~Andreas and G.~Curio,
  Phys.\ Lett.\  B {\bf 655} (2007) 290
  [arXiv:0706.1158 [hep-th]].
  
\bibitem{knw08}
  M.~Kuriyama, H.~Nakajima and T.~Watari,
  arXiv:0802.2584 [hep-ph].

\bibitem{lrx06}
  D.~Lust, S.~Reffert, E.~Scheidegger and S.~Stieberger,
  arXiv:hep-th/0609014.

\bibitem{gkx08}
  S.~Groot~Nibbelink, D.~Klevers, F.~Ploger, M.~Trapletti and 
  P.~K.~S.~Vaudrevange,
  JHEP {\bf 0804} (2008) 060
  [arXiv:0802.2809 [hep-th]].

\bibitem{dw08}
  R.~Donagi and M.~Wijnholt,
  arXiv:0802.2969 [hep-th].
  
\bibitem{bhv08}
  C.~Beasley, J.~J.~Heckman and C.~Vafa,
  arXiv:0802.3391 [hep-th];
  arXiv:0806.0102 [hep-th].

\bibitem{tw08}
  R.~Tatar and T.~Watari,
  arXiv:0806.0634 [hep-th].

\bibitem{drw82}
  S.~Dimopoulos, S.~Raby and F.~Wilczek,
  Phys.\ Lett.\  B {\bf 112} (1982) 133.

\bibitem{hn01}
L.~J. Hall and Y.~Nomura, Phys. Rev. \textbf{D64} (2001) 055003
  [hep-ph/0103125].

\bibitem{abc03}
  T.~Asaka, W.~Buchmuller and L.~Covi,
  Phys.\ Lett.\  B {\bf 563} (2003) 209
  [hep-ph/0304142].

\bibitem{bn02}
  G.~Burdman and Y.~Nomura,
  Nucl.\ Phys.\  B {\bf 656} (2003) 3
  [arXiv:hep-ph/0210257].

\bibitem{lee06}
  H.~M.~Lee,
  Phys.\ Lett.\  B {\bf 643} (2006) 136
  [arXiv:hep-ph/0609064].

\bibitem{drw08}
  B.~Dundee, S.~Raby and A.~Wingerter,
  arXiv:0805.4186 [hep-th].


\bibitem{akx08}
For a recent discussion and references, see\\
T.~Araki, T.~Kobayashi, J.~Kubo, S.~Ramos-Sanchez, M.~Ratz and 
P.~K.~S.~Vaudrevange,
  arXiv:0805.0207 [hep-th].


\end{thebibliography}
\end{document}